\documentclass{svproc}
\usepackage{graphicx}
\usepackage{caption}
\usepackage{subcaption}
\usepackage{amsmath}
\usepackage{comment}
\usepackage{float}
\usepackage{adjustbox}
\usepackage{multirow}
\usepackage{xcolor}

\begin{document}
\mainmatter              
\title{Higher-Order Temporal Network Prediction and Interpretation}
\titlerunning{Higher-Order Temporal Network Prediction}  
%
\author{H.A. Bart Peters \and Alberto Ceria \and Huijuan Wang}
\authorrunning{Peters et al.} 
%
\tocauthor{H.A. (Bart) Peters, Alberto Ceria, and Huijuan Wang}
\institute{Delft University of Technology, Mekelweg 4, 2628 CD Delft, Netherlands,\\
\email{H.Wang@tudelft.nl},\\
}

\maketitle              

\begin{abstract}
A social interaction (so-called higher-order event/interaction) can be regarded as the activation of the hyperlink among the corresponding individuals. Social interactions can be, thus, represented as higher-order temporal networks, that record the higher-order events occurring at each time step over time. The prediction of higher-order interactions is usually overlooked in traditional temporal network prediction methods, where a higher-order interaction is regarded as a set of pairwise interactions. The prediction of future higher-order interactions is crucial to forecast and mitigate the spread the information, epidemics and opinion on higher-order social contact networks. 
In this paper, we propose novel memory-based models for higher-order temporal network prediction. By using these models, we aim to predict the higher-order temporal network one time step ahead, based on the network observed in the past. Importantly, we also intent to understand what network properties and which types of previous interactions enable the prediction. The design and performance analysis of these models are supported by our analysis of the memory property of networks, e.g., similarity of the network and activity of a hyperlink over time respectively. Our models assume that a target hyperlink's future activity (active or not) depends the past activity of the target link and of all or selected types of hyperlinks that overlap with the target. We then compare the performance of both models with a baseline utilizing a pairwise temporal network prediction method. In eight real-world networks, we find that both models consistently outperform the baseline and the refined model tends to perform the best. Our models also reveal how past interactions of the target hyperlink and different types of hyperlinks that overlap with the target contribute to the prediction of the target's future activity.

\end{abstract}
\section{Introduction}
Temporal networks have been used to represent complex systems with time-varying network topology, where a link between two nodes is activated only when the node pair interacts \cite{holme2012temporal,masuda2016guide,holme2015modern}. This classic temporal network presentation assumes interactions to be pairwise. Social contacts/interactions have been mostly
studied as pairwise temporal networks. Social interactions, on the other hand, could be beyond pairwise, as individuals may interact in groups \cite{battiston2020networks,battiston2021physics,sekara2016fundamental}. For example, a collaboration in a scientific paper may engage more than two authors \cite{patania2017shape}. Such group interactions, involving an arbitrary number of nodes, are called higher-order interactions or events. Social contacts are thus better represented by higher-order temporal networks, consisting of higher-order interactions, i.e., the activation of hyperlinks (groups of nodes) at specific times.\\

\noindent The classic temporal network prediction problem aims to predict pairwise contacts, one time step ahead, based on the network observed in the past. This problem has applications in a variety of research fields. Predicting a temporal network in the future enables better forecast and mitigation of the spread of epidemics or misinformation on the network. This prediction problem is also equivalent to problems in recommender systems: e.g., predicting which user will purchase which product, or which individuals will become acquaintances at the next time step \cite{lu2012recommender,aleta2020link}. \\

\noindent Methods have been proposed for pairwise temporal network prediction. Some rely on network embeddings, where nodes are represented as points in a low dimensional embedding space. Within this space, connected nodes are then supposed to be close \cite{zhou2019dynamic}. Alternatively, deep learning methods have been proposed \cite{li2014deep}. Examples are adversarial networks \cite{chen2019generative} or LSTM methods \cite{chen2019lstm}. The downside of deep learning methods, however, is that they are at the expense of high computational costs, and are limited in providing insights regarding which network mechanisms enable the network prediction. Recently, Zou et al. \cite{zou2023memory} observed time-decaying memory in pairwise temporal networks. This means that snapshots of the network closer in time share more similarities. This observation was then used in the design a network-based prediction method. Lastly, other methods have been proposed to predict whether a set of nodes will have at least one group interaction in the future \cite{benson2018simplicial,liu2023higher,piaggesi2022effective} and when the first group interaction among these nodes will occur \cite{liu2022neural}. \\

\noindent The forecast of hyperlink activity is usually overlooked in these traditional prediction methods. In this paper, we aim to predict higher-order interactions, one time step ahead, based on the higher-order temporal network observed in the past for a duration $L$. Moreover, we intent to understand what network properties and which types of previous interactions enable the prediction. 
Recent results have shown that, in physical contact networks, the activity of hyperlinks with different orders overlapping in component nodes seems temporally correlated \cite{cencetti2021temporal,gallo2023higher,ceria2023temporal}, likely as a result of the dynamics of splitting or merging of social groups \cite{iacopini2023temporal}. Motivated by this observation and the time-decaying memory found in pair-wise temporal networks, we first explore the memory property, or similarity over time, of higher-order temporal networks. Specifically, we study systematically to what extent a network's topology, the activity of a hyperlink and the activity between neighboring hyperlinks, respectively, remains similar over time. 
After this analysis, we propose a memory-based model to solve the prediction problem utilizing the memory property observed. This model is a generalization to higher-order of the pairwise model proposed in \cite{zou2023memory}. \\

\noindent Our model assumes that the activity (interacting or not) of a hyperlink (or group) at the next time step is influenced by the past activity of this target hyperlink and all neighboring hyperlinks. The time-decaying memory observed is then incorporated in our model through a universal exponential decay, which ensures that the activity of recent events is more influential than the activity of older events. In the prediction problem, we assume that the groups, each of which interact at least once in the network (in the past or future), as well as the total number of events per group size (order) at the prediction time step are known. These assumptions aim to simplify the problem. Beyond, the total number of interactions of each order could be influenced by factors like weather and holidays, other than the network observed in the past. This assumption means that group friendship is known, and we aim to predict which groups with group friendship interact at the prediction step. \\

\noindent Using this generalized model, we unravel the contribution of the activity of each type of neighboring hyperlinks when forecasting a target hyperlink's future activity. This together with the memory property observed between neighboring hyperlinks, motivates us to propose a refined model upon the generalized model. The refined model utilizes the activity of more relevant types of neighboring links: sub-hyperlinks (sub-groups of the target group) and super-hyperlinks (super-groups in which the target group is included), as well as the target hyperlink. We also propose a baseline model that uses a memory-based pairwise temporal network prediction method \cite{zou2023memory}: it considers the higher-order temporal network observed in the past as a pairwise temporal network, then predicts the pairwise temporal network in the next time step and ultimately deduces higher-order interactions from the predicted pairwise interactions at the same prediction time step. In the end, we find that both our models generally perform better than the baseline, as evaluated in eight real-world physical contact networks. Moreover, the refined model outperforms the generalized model for the prediction of events of order $2$ and $3$. Lastly, we find that the past activity of the target group is the most important factor in the prediction, followed by the activity of the target group's super- and sub-groups. \\

\noindent This paper is an extension of our previous work \cite{jung-muller2024higher}, which considered only the refined model without the systematic analysis of memory property nor the generalized model to motivate the design of the refined model.

\section{Network representation} \label{section-definitions}
Before stating the definition of a higher-order temporal network, we will start with the definition of the pairwise case. A pairwise temporal network \(G\) measured at discrete times can be represented as a sequence of network snapshots \( G= \{G_1, G_2, ..., G_T \}  \), where \(T\) is the duration of the observation window, \( G_t = (V, E_t) \) is the snapshot at time step \(t\). Here \(V\) and \(E_t\) are the set of nodes and interactions, respectively. If nodes \(a\) and \(b\) have a contact at time step \(t\), then \( (a, b) \in E_t\). Here, we assume that all snapshots share the same set of nodes \(V\). The set of links in the time-aggregated network is defined as \( E = \bigcup_{t=1}^{t=T} E_t \). In this aggregated network, a pair of nodes is connected with a link if at least one contact occurs between them in the temporal network. The temporal connection or activity of link $i$ over time can be represented by a $T$-dimensional vector $x_i$. The elements of this vector are given by $x_i(t)$, where \( t \in [1, T]\). We have \(x_i(t) = 1\) when link \(i\) has a contact at time step \(t\), and \(x_i(t) = 0\) if no contact occurs at \(t\). A temporal network can thus be equivalently represented by its aggregated network, where each link \(i\) is further associated with its activity time series \(x_i\).\\

\noindent Social interactions, which may involve more than two individuals, can be more accurately represented by the activation/interaction of hyperlinks in a higher-order temporal network \(H\). This network is a sequence of network snapshots \(H = \{H_1, ..., H_T \}\), where \( H_t = (V, {\mathcal{E}}_t) \) is the snapshot at time step \(t\). Here, \(V\) represents the set of nodes shared by all snapshots and \( {\mathcal{E}_t}\) is the set of hyperlinks that are activated at time step \(t\). The activation of a hyperlink \( (u_1, ..., u_d) \) then corresponds to a group interaction among nodes \(u_1\), ..., \(u_d\). The hyperlink \( (u_1, ..., u_d) \) active at time step \(t\) is called an event or interaction of size \(d\).
The set of hyperlinks in the higher-order time-aggregated network is defined as \( \mathcal{E} = \bigcup_{t=1}^{t=T} {\mathcal{E}}_t \). A hyperlink belongs to \(\mathcal{E}\) if it is activated at least once in the temporal network.
A higher-order temporal network can thus be equivalently represented by its higher-order aggregated network, where each hyperlink \(i\) is further associated with its activity time series \(x_i\).\\

\noindent The previous activity of neighboring hyperlinks may contribute to the prediction of a target hyperlink's activity in the future. We therefore define different types of neighboring hyperlinks for a given target hyperlink. For an arbitrary target hyperlink $h$, an arbitrary neighboring hyperlink $h'$ is called a type $\phi = (dd'o)$-neighboring hyperlink of $h$. Here, $d$ and $d'$ are the sizes of hyperlinks $h$ and $h'$, respectively, while $o$ is the number of nodes shared between both hyperlinks. It holds that $1 \leq o \leq \min(d, d')$. The unique $\phi = (dd'o)$-neighboring hyperlink when $d = d' = o$ refers to the target hyperlink $h$ itself. Two other particular classes of $\phi$-neighboring hyperlinks are the so-called sub- and super-hyperlinks. We say that $h'$ is a sub-hyperlink of $h$ if $h' \subset h$, i.e., if $h'$ is included in $h$. At the same time, we call $h$ a super-hyperlink of $h'$. Given an arbitrary hyperlink of order $d$, all types $\phi$ of neighboring hyperlinks reside in the set $\Phi^d$. Table \ref{tab:PHIset} displays all possible values of $\phi\in\Phi^d$ for each target group of order $d \in [2,3,4]$.


\begin{table}[ht!]
    \begin{adjustbox}{center}
    \begin{tabular}{|c|c|}
        \hline
        order $d$ & $\phi$  \\ \hline
        $2$ & $\mathbf{222},\;221,\;231,\;\mathbin{\color{red}232},\;241,\;\mathbin{\color{red}242}$\\ \hline
        $3$ & $\mathbf{333},\;321,\;\mathbin{\color{blue}322},\;331,\;332,\;341,\;342,\;\mathbin{\color{red}343}$\\ \hline
        $4$ & $\mathbf{444},\;421,\;\mathbin{\color{blue}422},\;431,\;432,\;\mathbin{\color{blue}433},\;441,\;442,\;443$\\ \hline
    \end{tabular}
     \end{adjustbox}
\caption{All possible types $\phi\in\Phi^d$ of neighboring hyperlinks for an order $d$ target hyperlink.  The $\phi$-values corresponding to the target, sub- and super-hyperlinks are displayed in bold, blue and red, respectively. }
\label{tab:PHIset}
\end{table}

\section{Datasets} \label{section-datasets}

To design and evaluate our network prediction methods, we consider eight empirical physical contact networks from the SocioPatterns project (sociopatterns.org). They are collections of face-to-face interactions in social contexts such as study places (Highschool2012 \cite{fournet2014contact}, Highschool2013 \cite{mastrandrea2015contact}, Primaryschool \cite{stehle2011high}), conferences (SFHH Conference \cite{genois2018can}, Hypertext2009 \cite{isella2011s}), workplaces (Hospital \cite{vanhems2013estimating}, Workplace \cite{genois2018can}) or an art gallery (Science Gallery \cite{isella2011s}). These events are recorded as a set of pairwise interactions, provided that the distance between people is smaller than $2 m$. Based on these pairwise contacts, group interactions are then deduced by promoting every fully-connected clique of \(\binom{d}{2}\) contacts to an event of size \(d\). Since a clique of order \(d\) contains all its sub-cliques of orders \(d' < d \), only the maximal clique is promoted to a higher-order event. This method has been used in \cite{cencetti2021temporal} and \cite{ceria2023temporal}, where sub-cliques are ignored as well. \\

\begin{figure}[h!]
    \centering
    \includegraphics[width=.90\textwidth]{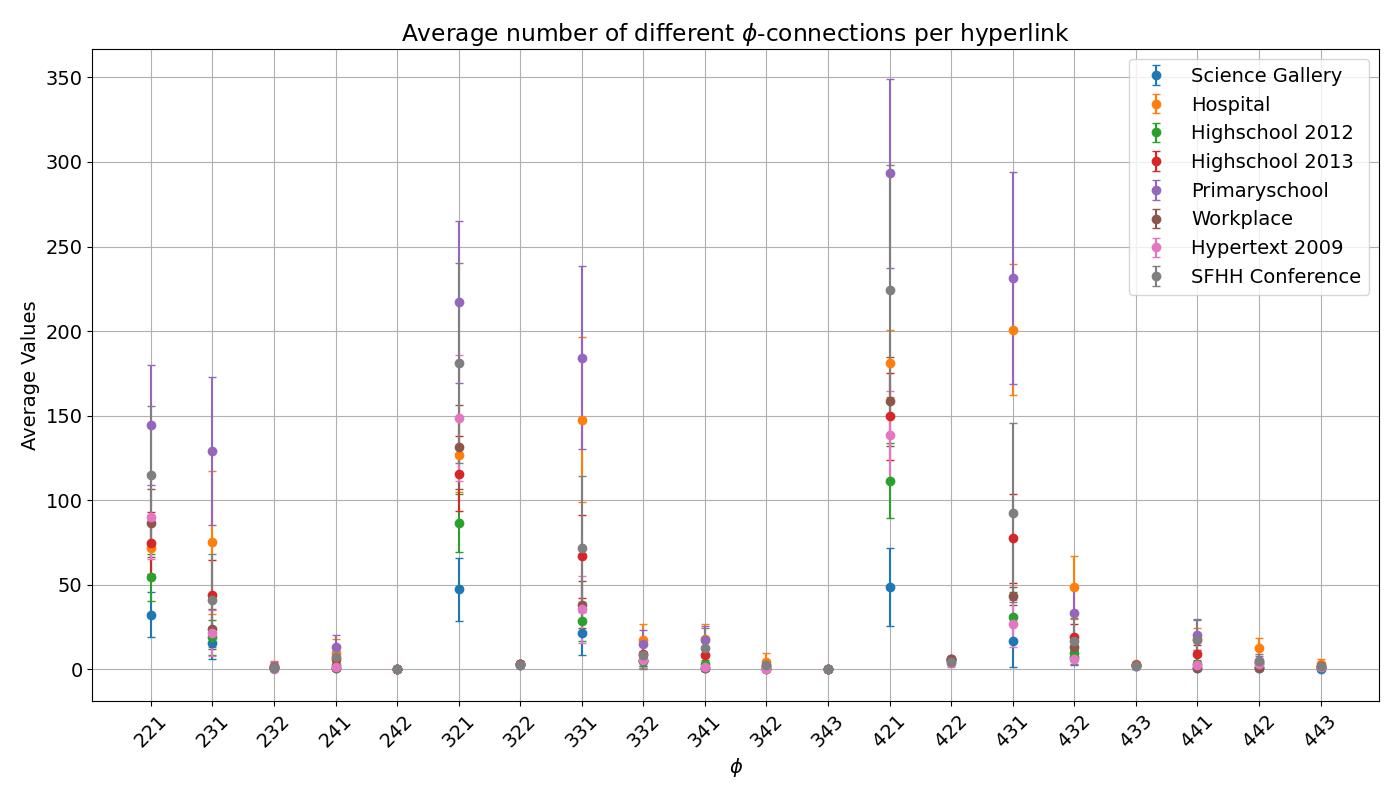}
    \caption{Average number of $\phi$-neighbors per hyperlink, including error bars, for every dataset.}
    \label{fig:phi_avg_per_hyperlink}
\end{figure}

\noindent The datasets are further preprocessed in a similar way as done in \cite{ceria2022topological} and \cite{ceria2023temporal}. We start by removing nodes that are not connected to the largest connected component in the pairwise time-aggregated network. After that, we neglect long periods of inactivity in the network. Such periods usually correspond to nights and weekends, and are therefore recognized as outliers. We then end up with eight preprocessed datasets, whose total number of events of each order is shown in Table \ref{tab:dataset_table}. Figure \ref{fig:phi_avg_per_hyperlink} displays the average number of neighboring hyperlinks of each type  $\phi$ per target hyperlink of a given order $d$, in each higher-order temporal network. In general, we find that an order $d$ hyperlink has on average evidently more type $\phi_1$-neighbors than $\phi_2$-neighbors, if $\phi_1 = (dd'o_1)$, $\phi_2 = (dd'o_2)$ and $o_1<o_2$. In other words, there are more order $d'$ neighboring hyperlinks that overlap less with the target link than order $d'$ neighbors that overlap more with the target. In the next section, we will explore the correlation between a target's activity and the activity of each type $\phi$ of neighboring hyperlink, either overlapping significantly with the target or not. 


\begin{table}[ht!]
    \begin{adjustbox}{center}
    \begin{tabular}{|c|c|c|c|c|c|}
        \hline
        Dataset & Order 2 & Order 3 & Order 4 & Order 5 & Order 6+ \\
        \hline\hline
        Science Gallery & 12770 & 1421 & 77 & 7 & 0 \\ \hline
        Hospital & 25487 & 2265 & 81 & 2 & 0 \\ \hline
        Highschool 2012 & 40671 & 1339 & 91 & 4 & 0 \\  \hline
        Highschool 2013 & 163973 & 7475 & 576 & 7 & 0 \\ \hline
        Primaryschool & 97132 & 9262 & 471 & 12 & 0 \\ \hline
        Workplace & 71529 & 2277 & 14 & 0 & 0 \\ \hline
        Hypertext 2009 & 18120 & 874 & 31 & 12 & 4 \\ \hline
        SFHH Conference & 48175 & 5057 & 617 & 457 & 199 \\ \hline 
    \end{tabular}
     \end{adjustbox}
\caption{ Number of events of every order for every dataset after preprocessing. }
\label{tab:dataset_table}
\end{table}

\section{Network memory property} \label{section-jaccard}
In this section,  we aim to explore the memory property of higher-order temporal networks, which will be utilized in the model design to predict higher-order events. We will first examine the Jaccard similarity between network snapshots and the auto-correlation of a hyperlink's activity time series. Both similarity measures will then be studied as a function of time lag between snapshots or time series, respectively. They will give us insights in the memory/similarity in network topology and activity of each hyperlink overtime, respectively. After that, we will explore if the activity of a target hyperlink/group is correlated with the activity of each type of neighboring hyperlinks. We will also examine this correlation as a function of time lag, thereby studying if certain types of neighboring links poss memory of the target link in activity. 

\subsection{Jaccard similarity between network snapshots}
Firstly, we explore the similarity of a higher-order temporal network at order $d$, which contains only order $d$ events in the network, over time. We examine the Jaccard similarity $J_d(\Delta)$ of a higher-order temporal network at order $d$ at two different time steps, separated by a time lag $\Delta$, where $d \in [2, 5]$. The Jaccard similarity measures how similar two given sets are by taking the ratio of the size of their intersection set over the size of their union set. In our case, we compute the Jaccard similarity 
\begin{equation}
    \label{eq:Jaccard}
    \frac{ |\mathcal{E}_{t}^d \cap \mathcal{E}_{t+\Delta}^d| }{ |\mathcal{E}_{t}^d \cup \mathcal{E}_{t+\Delta}^d| }
\end{equation}
between  \( \mathcal{E}_{t}^d \), the set of \(d\)-hyperlinks (hyperlinks of order \(d\)) active at time step \(t\), and \( \mathcal{E}_{t+\Delta}^d \). Its average over time $t$ is referred to as the Jaccard similarity $J_d(\Delta)$ of a higher-order temporal network at order $d$ with a time lag $\Delta$. A large $J_d(\Delta)$ implies a large overlap/similarity between two
snapshots of a temporal network at order $d$ with a time lag $\Delta$. \\

\begin{figure*}[h!]
    \centering
    \begin{subfigure}[b]{0.45\textwidth}
        \includegraphics[width=\textwidth]{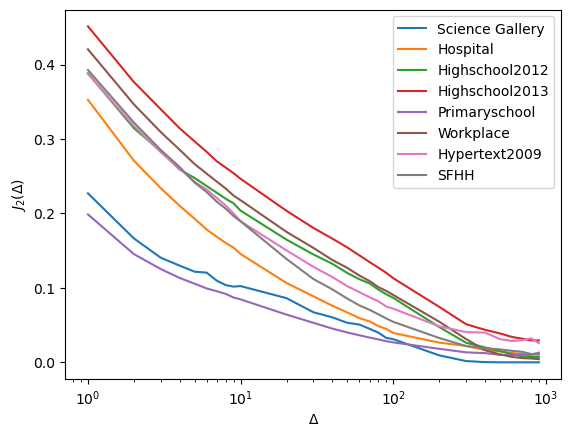}
        \caption{$d = 2$}
        \label{fig:jaccard_2}
    \end{subfigure}
    \begin{subfigure}[b]{0.45\textwidth}
        \includegraphics[width=\textwidth]{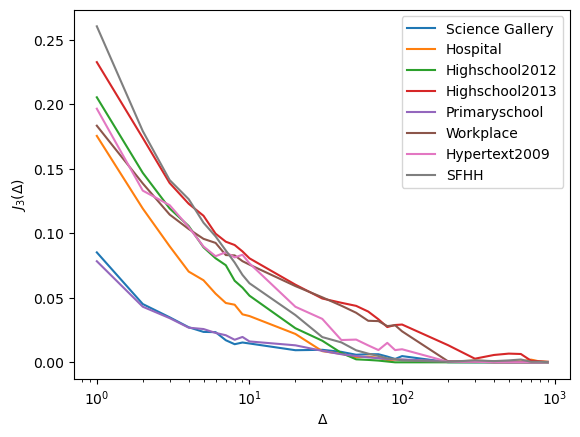}
        \caption{$d = 3$}
        \label{fig:jaccard_3}
    \end{subfigure}
    \begin{subfigure}[b]{0.45\textwidth}
        \includegraphics[width=\textwidth]{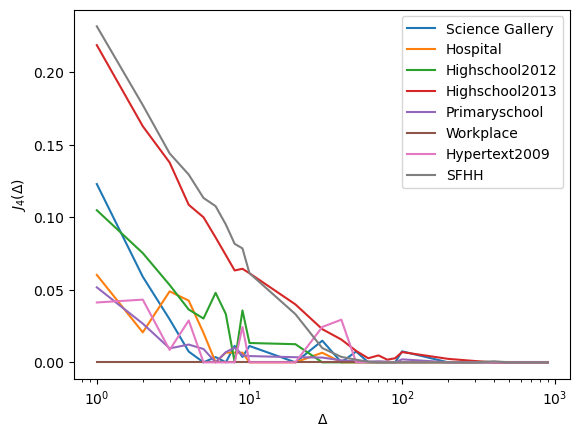}
        \caption{$d = 4$}
        \label{fig:jaccard_4}
    \end{subfigure}
    \begin{subfigure}[b]{0.45\textwidth}
        \includegraphics[width=\textwidth]{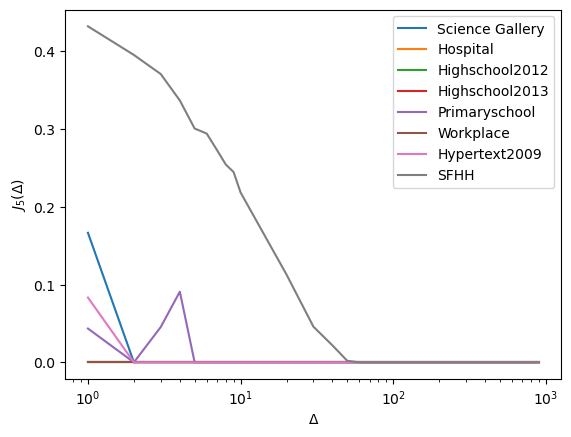}
        \caption{$d = 5$}
        \label{fig:jaccard_5}
    \end{subfigure}
    \caption{Jaccard similarities $J_d(\Delta)$ of eight real-world physical contact networks for events of order $d \in [2,5]$ as a function of the time lag $\Delta$.}
    \label{fig:jaccard}
\end{figure*}

\noindent Figure \ref{fig:jaccard} displays the Jaccard similarity of each real-world higher-order temporal network at each order $d$. We observe that, for all datasets, the similarity decays as the time lag increases at each order $2$, $3$ and $4$. This time-decaying memory at order $5$ is only observed in the SFHH network, which is the only network that has a non-negligible number of order $5$ events, as shown in Table \ref{tab:dataset_table}. Besides that, we find that as the order increases, the slope of the curves increases accordingly.
\subsection{Auto-correlation of the activity of a hyperlink}
\begin{figure*}[h!]
    \centering
    \begin{subfigure}[b]{0.45\textwidth}
        \includegraphics[width=\textwidth]{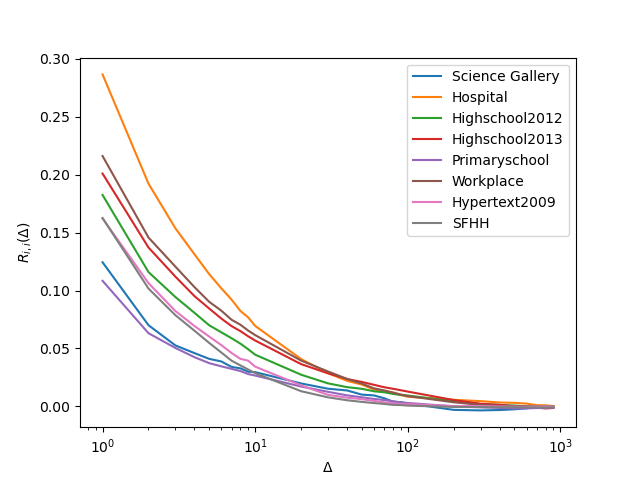}
        \caption{$d = 2$}
        \label{fig:autocor_2}
    \end{subfigure}
    \begin{subfigure}[b]{0.45\textwidth}
        \includegraphics[width=\textwidth]{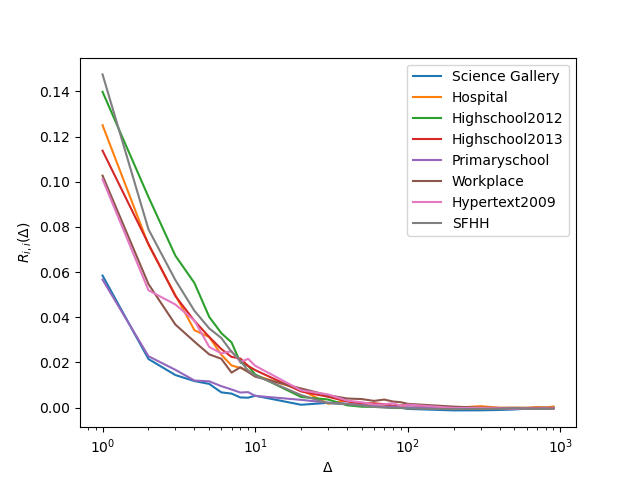}
        \caption{$d = 3$}
        \label{fig:autocor_3}
    \end{subfigure}
    \begin{subfigure}[b]{0.45\textwidth}
        \includegraphics[width=\textwidth]{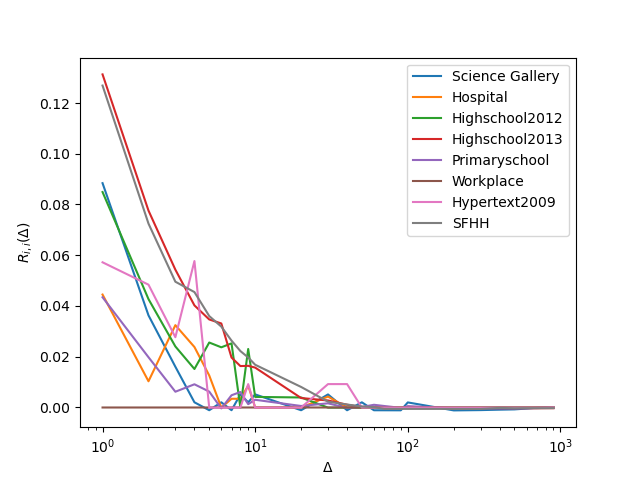}
        \caption{$d = 4$}
        \label{fig:autocor_4}
    \end{subfigure}
    \begin{subfigure}[b]{0.45\textwidth}
        \includegraphics[width=\textwidth]{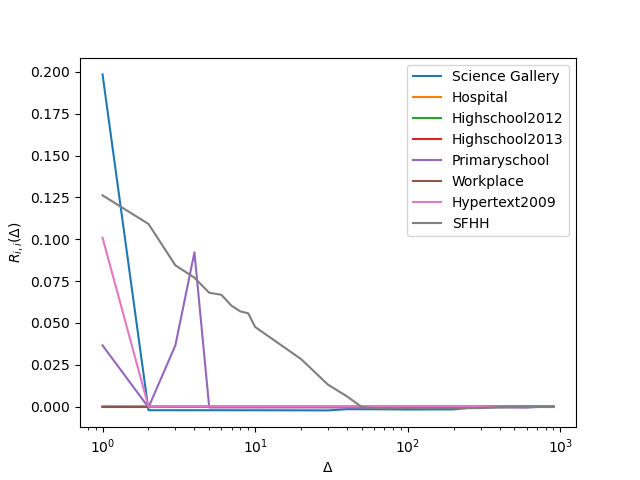}
        \caption{$d = 5$}
        \label{fig:autocor_5}
    \end{subfigure}
    \caption{The average auto-correlations $R_{i,i}(\Delta)$ of the activity of an order $d$ hyperlink as a function of the time lag $\Delta$ in each real-world physical contact network.}
    \label{fig:autocor}
\end{figure*}

Secondly, we look at the average auto-correlation of the activity time series of a hyperlink in a higher-order aggregated network. The auto-correlation of a time series is the Pearson correlation between the given time series and its lagged version. Concretely, the auto-correlation of the activity of a hyperlink $i$ is the Pearson correlation coefficient $R_{i,i}(\Delta)$ between $\{x_i(t)\}_{t = 1, 2, ..., T-\Delta}$ and $\{x_i(t)\}_{t = \Delta+1, \Delta+2, ..., T}$. Figure \ref{fig:autocor} shows the average auto-correlation $R_{i,i}(\Delta)$ over all hyperlinks of order $d$, where $d \in [2,5]$. Similar to the Jaccard similarities displayed in Figure \ref{fig:jaccard}, we observe time-decaying memory or auto-correlation for the activity of each hyperlink of any order $2$, $3$ or $4$, and the decay is faster if the order $d$ is larger.

\noindent The Jaccard similarity observed at network level for each order, together with the auto-correlation observed at each hyperlink, indicate that the previous activity of a hyperlink could be useful for the prediction of the future activity of this hyperlink.

\subsection{Correlation between neighboring hyperlinks in activity} \label{section_phi_memory}
Figure \ref{fig:phi_avg_per_hyperlink} shows the average number of each type $\phi$-neighboring hyperlinks possessed by each target hyperlink in each dataset. We wonder whether a target link shares similarity in activity with its neighboring links, such that we could use the previous activity of neighboring links to predict the future activity of a given target link.  In order to study this, we first compute, for each target hyperlink $i$, the Pearson correlation coefficient $R_{i,\phi}(\Delta)$ between the activity of the target $i$ and the sum of the activity time series of all its $\phi$-neighboring hyperlinks with a time lag $\Delta$. \\

\begin{figure*}[h!]
    \centering
    \begin{subfigure}[b]{0.4\textwidth}
        \includegraphics[width=\textwidth]{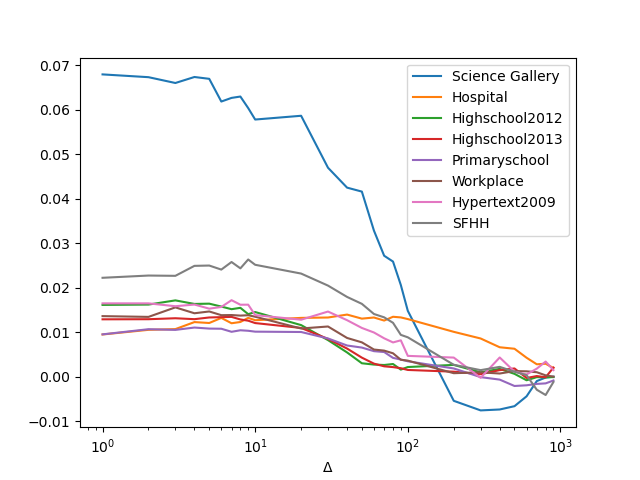}
        \caption{$\phi = 321$}
        \label{fig:pearsoncor_321}
    \end{subfigure}
    \begin{subfigure}[b]{0.4\textwidth}
        \includegraphics[width=\textwidth]{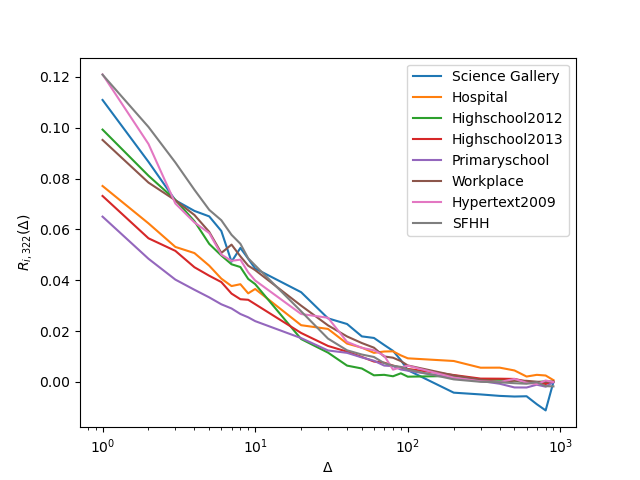}
        \caption{$\phi = 322$}
        \label{fig:pearsoncor_322}
    \end{subfigure}
    \begin{subfigure}[b]{0.4\textwidth}
        \includegraphics[width=\textwidth]{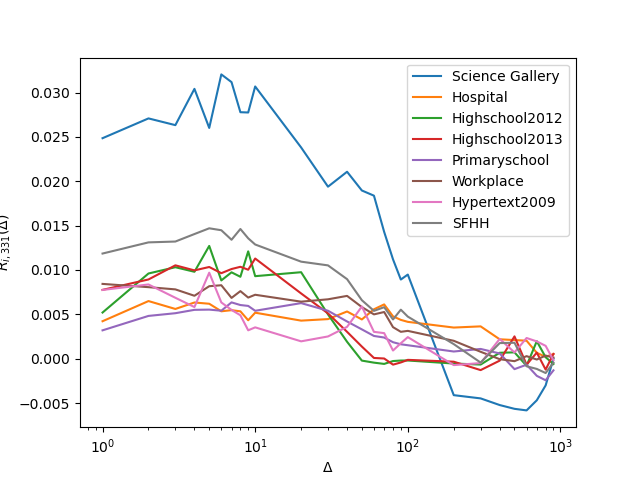}
        \caption{$\phi = 331$}
        \label{fig:pearsoncor_331}
    \end{subfigure}
    \begin{subfigure}[b]{0.4\textwidth}
        \includegraphics[width=\textwidth]{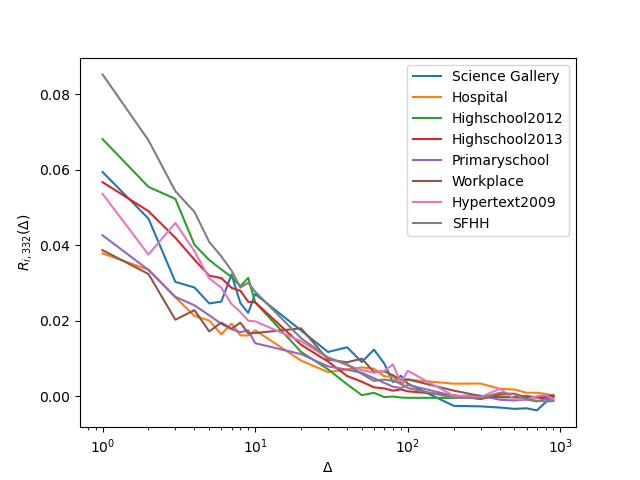}
        \caption{$\phi = 332$}
        \label{fig:pearsoncor_332}
    \end{subfigure}
    \begin{subfigure}[b]{0.4\textwidth}
        \includegraphics[width=\textwidth]{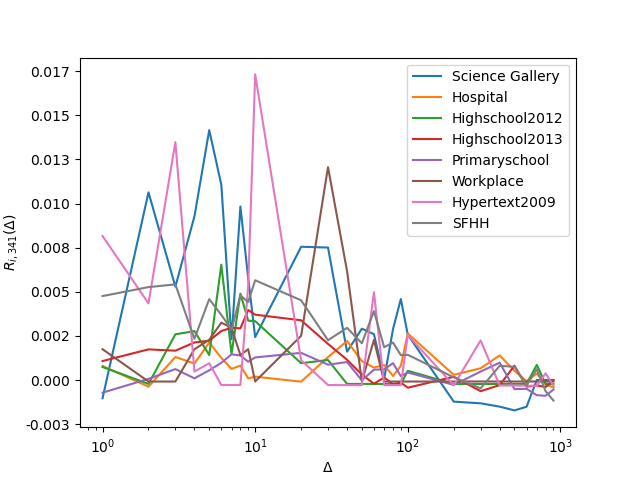}
        \caption{$\phi = 341$}
        \label{fig:pearsoncor_341}
    \end{subfigure}
    \begin{subfigure}[b]{0.4\textwidth}
        \includegraphics[width=\textwidth]{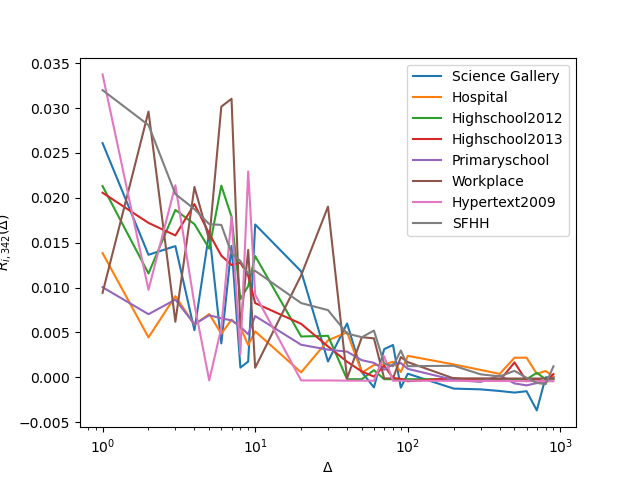}
        \caption{$\phi = 342$}
        \label{fig:pearsoncor_342}
    \end{subfigure}
    \begin{subfigure}[b]{0.4\textwidth}
        \includegraphics[width=\textwidth]{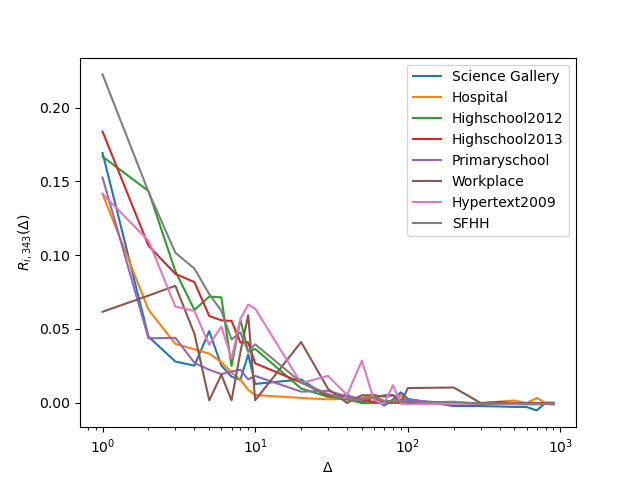}
        \caption{$\phi = 343$}
        \label{fig:pearsoncor_343}
    \end{subfigure}
    \caption{Average Pearson correlation coefficient $R_{i,\phi}(\Delta)$ for order $3$ hyperlinks in eight real-world physical contact networks as a function of time lag $\Delta$.}
    \label{fig:pearsoncor_rest3}
\end{figure*}

\noindent We take order $3$ target hyperlinks as an example. Figure \ref{fig:pearsoncor_rest3} shows the average Pearson correlation coefficient $R_{i,\phi}(\Delta)$ over order $3$ target hyperlinks that have at least a $\phi$-neighbor. We find that the correlation is more evident with type $\phi=322$ and $\phi=343$ neighbors, which are sub- and super-hyperlinks of the target hyperlink, respectively, and with $\phi=332$-neighbors. The correlation with other types of neighbors is negligibly small, though they are more present in all datasets as shown in Figure \ref{fig:phi_avg_per_hyperlink}. The correlation with these three types of neighbors respectively is also decaying with $\Delta$. \\



\noindent The average Pearson correlation coefficients $R_{i,\phi}(\Delta)$, between an order $2$ or $4$ target hyperlink and a $\phi$-neighboring hyperlink with time lag $\Delta$, is displayed in Figures \ref{fig:pearsoncor_rest2} and \ref{fig:pearsoncor_rest4} in the appendix, respectively. The same has been observed that the correlation in activity between a target link with a sub- or super-hyperlinks tends to be stronger, and this correlation/memory is time-decaying. Moreover, among the super- and sub-hyperlinks of a target hyperlink of any order, those that overlap more with the target in nodes are more strongly correlated with the target link.


\section{Models} \label{section-models}
\subsection{Baseline}
We propose a baseline for higher-order temporal network prediction utilizing the following pairwise network prediction model, called Self-Driven SD model, proposed in \cite{zou2023memory}. The SD model is a memory-based model that predicts a pairwise link's activity at the next time step, based on its past activity. The SD model estimates the tendency \( w_i(t+1) \) for each link \(i\) to be active at time \(t+1\) as
\begin{equation}
    \label{eq:SD}
    w_i(t+1)= \sum^{k=t}_{k=t-L+1} x_i(k) e^{-\tau(t-k)}.
\end{equation}
Here, \(t+1\) is the prediction time step, \(L\) is the length of the past observation used for the prediction, \(\tau\) is the exponential decay factor and \(x_i(k)\) is the activation state of link \(i\) at time \(k\). So we have \( x_i(k)=1 \) if the link \(i\) is active at time \(k\) and \( x_i(k)=0 \) otherwise. The exponential function in Equation (\ref{eq:SD}) captures the memory of the network, thereby ensuring that recent events have more influence than older events. We compute the activation tendency for each link in the pairwise aggregated network at the prediction step \(t+1\). Given the pairwise aggregated network and the number of pairwise contacts occurring at \(t+1\), the same number of links with the highest activation tendency at \(t+1\) will then be predicted to be active. \\

\noindent The baseline model is constructed as follows. First, we consider the higher-order temporal network observed for the past $L$ time steps as a pairwise temporal network. After that, we apply the SD model to predict the pairwise interactions at the prediction step. From those interactions, we then deduce higher-order interactions using the same method that promotes pairwise interactions that form a clique to a higher-order event as described in Section \ref{section-datasets}. This set of higher-order interactions is considered the prediction made by the baseline model.

\subsection{Generalized model} \label{section_generalized_model}
Beyond pairwise temporal networks, we have also observed time-decaying memory in higher-order temporal networks at each group order. This motivates us to generalize the SD model for higher-order network prediction. The essence of this generalized model is that the future activity of a hyperlink should be dependent on its past activity.
Furthermore, Figure \ref{fig:pearsoncor_rest3} showed us that the activity of a target hyperlink shares similarity with the activity of its neighboring hyperlinks. Hence, the activity of a hyperlink is possibly dependent on the past activity of its neighboring hyperlinks. Finally, recent events should have more influence than older events, based on the time-decaying memory observed in the auto-correlation of the activity of each hyperlink and the correlation in activity between a target and its neighboring hyperlinks. Therefore, we propose that the activation tendency of a hyperlink $i$ at a prediction time step $t+1$ is given as:
\begin{equation}
    \label{eq:generalmodal}
    w_i(t+1) = \sum_{\phi \in \Phi^{d_i}} c_\phi y_i^\phi(t) + c_{d_i},
\end{equation}
where
\begin{equation}
    \label{eq:xphi}
    y_i^\phi(t) = \sum^{k=t}_{k=t-L+1} \; \sum_{j \in S_i^\phi} x_j(k) e^{-\tau(t-k)}.
\end{equation}
In Equation (\ref{eq:xphi}), $L$ resembles the time span used for prediction, $\tau$ is the exponential decay factor, and $S_i^\phi$ corresponds to set of all type $\phi$ neighboring hyperlinks of the target $i$. The contribution of each type $\phi$ of neighboring hyperlinks $y_i^\phi(t)$, to estimate the activation tendency $w_i(t+1)$ of a target hyperlink $i$, includes the activity of all hyperlinks in the set $S_i^\phi$, over the observation period of length $L$ with an exponential decay over time. For a target hyperlink $i$ of order $d_i$, the contribution of all types $\phi\in\Phi^{d_i}$ of possible neighboring hyperlinks is taken into account. The activation tendency $w_i(t+1)$ is assumed to be a linear function of the contribution of each type $\phi \in \Phi^{d_i}$ of neighboring hyperlinks associated with a coefficient $c_\phi$, which we assume to be the same for all target hyperlinks of the same order. \\

\noindent All coefficients will be learned via Lasso regression that minimizes the objective function
\begin{equation}
    \label{eq:Lasso}
    \min_{C^d} \left\{ \sum_i \sum_{t+1} \left(( x_i(t+1) - w_i(t+1) \right)^2 + \alpha \sum_{c\in C^d} |c| \right\},
\end{equation}
where $x_i(t+1)$ is the true activity of hyperlink $i$ at $t+1$, and $C^d$ is the set consisting of the intercept $c_d$ and all coefficients $c_\phi$, where $\phi\in\Phi^d$ for a target group of order $d$. A unique set of coefficients will be learned for all target hyperlinks with a given order $d$, by considering the activity of each order $d$ hyperlink $i$ at every time step $t+1 \in [L+1,T]$ as the target to predict thus test set and the corresponding activity of this link $i$ and its all types of neighboring hyperlinks in the previous $L$ time steps as the training set. \\

\noindent We use L1 regularization, which adds a penalty to the sum of the magnitude of coefficients. The parameter $\alpha$ controls the penalty strength. The regularization forces some of the coefficients to be zero and thus leads to models with few non-zero coefficients (relevant features). The optimal $\alpha$ is achieved through a grid search in a set of $200$ logarithmically spaced points within $[10^{-19}, 10]$. \\

\noindent Using the learned optimal fitting coefficients, the activation tendency is computed for each hyperlink in the higher-order aggregated network, which is given in the prediction problem. Given the number \(n_{t+1}^{d}\) of events of each order \(d\) at the prediction step $t+1$, the \(n_{t+1}^{d}\) hyperlinks of order \(d\) with the highest activation tendency at \(t+1\) are then predicted to be active. Since the hyperlink forecast is performed for each order separately, the prediction of events of a given order does not influence the prediction of another order. \\

\subsection{Refined model}

Some types of neighboring hyperlinks are possibly more relevant for the prediction than the others. If so, better prediction can possibly be achieved via including only more relevant types of neighboring links instead of considering all types of neighboring links as in our generalized model. Therefore, we will explore the optimal coefficients learned for the generalized model discussed in Section \ref{section_generalized_model}, as well as the correlation in activity between neighboring hyperlinks analyzed in Section \ref{section_phi_memory}, to identify the relatively more relevant types of neighboring links and propose a refined model. \\

\noindent  
We consider $\tau=5$ and $L=30$ as an example for the model design and performance analysis. It has been shown that $\tau\in[0.5,5]$ leads to an optimal prediction accuracy for pairwise temporal network prediction \cite{zou2023memory}, thus there is no need to calibrate $\tau$. The choice of $L$ could be limited thus driven by the context of real-world prediction problem, e.g., how long of the network in the past can be used for the prediction. A small $L$ like $L=30$ is more feasible in practice and leads to a lower computational complexity. In Section \ref{section_performance_analysis_refined}, the influence of these two parameters on the prediction quality will be discussed. \\

\noindent The coefficients that weigh the contributions of all possible types of neighboring hyperlinks in estimating the connection tendency of an order $2$ and $3$ target hyperlink, obtained by Lasso regression, are given in Table \ref{tab:coeffs2} and \ref{tab:coeffs3}, respectively. 
We notice that the coefficients for sub- and super- hyperlinks tend to be larger than those for other types of neighboring hyperlinks. This is in line with our analysis in Section \ref{section_phi_memory}, where we showed that these neighboring groups have generally a larger correlation with target hyperlinks than other $\phi$-neighbors. Furthermore, the coefficient corresponding to the target hyperlink is the largest. This is likely because a) the auto-correlation or memory in the activity of a hyperlink itself, as shown in Figure \ref{fig:autocor}, is evident and b) although a target hyperlink could also be evidently correlated with a sub- or super-hyperlink in activity, many target links may not have any such sub- nor super-hyperlink, as supported by Figure \ref{fig:phi_avg_per_hyperlink}. These two observations regarding the coefficients are not always true for the coefficients used to predict order $4$ events, as shown in Table \ref{tab:coeffs4} in the appendix, due to the low number of order $4$ hyperlinks. \\

\noindent Both observations in coefficients when predicting order $2$ and $3$ events, supported by our network memory analysis, motivate us to consider the following refined model. It is the same as the generalized model, but includes only the contribution of the target hyperlink itself and the sub- and super- neighboring hyperlinks when estimating the connection tendency $w_i(t+1)$ of the target hyperlink. Similarly, the coefficients weighing the contributions of sub- and super-links could be learned via Lasso Regression. Using these optimal coefficients learned from the whole data set over $[1,T]$, the connection tendency of each target hyperlink can be computed at any prediction time $t+1 \in [L+1,T]$, based on the network observed within $[t-L+1, t]$. Given the total number of events of each order at $t+1$, the higher-order temporal network can be predicted by considering hyperlinks with the highest connection tendency at each order to be active. 



\begin{table}[h]
\begin{adjustbox}{center}
    \begin{tabular}{|c|c|c|c|c|c|c|c|}
        \hline
        Dataset & $c_{222}$ & $c_{221}$ & $c_{231}$ & $c_{232}$ & $c_{241}$ & $c_{242}$ & $c_2$  \\
        \hline\hline
        Science Gallery & 0.31 & 0.01 & 0.01 & 0.17 & 0.00 & 0.08 & 0.00  \\ \hline
        Hospital & 0.50 & 0.00 & 0.00 & 0.22 & 0.00 & 0.12 & 0.00  \\ \hline
        Highschool2012 & 0.56 & 0.00 & 0.00 & 0.20 & 0.01 & 0.08 & 0.00  \\ \hline
        Highschool2013 & 0.61 & 0.00 & 0.00 & 0.20 & 0.00 & 0.06 &  0.00 \\ \hline
        Primaryschool & 0.31 & 0.00 & 0.00 & 0.17 & 0.00 & 0.11 & 0.00 \\ \hline
        Workplace & 0.55 & 0.00 & 0.00 & 0.23 & 0.00 & 0.16 & 0.00 \\ \hline
        Hypertext2009 & 0.48 & 0.00 & 0.00 & 0.23 & 0.00 & 0.17 & 0.00  \\ \hline
        SFHH Conference & 0.52 & 0.00 & 0.00 & 0.19 & 0.00 & 0.04 & 0.00 \\ \hline
    \end{tabular}
    \end{adjustbox}
\caption{Coefficients of the generalized model to predict order $2$ events, when $L=30$ and $\tau=5$. }
\label{tab:coeffs2}
\end{table}

\begin{table}[h]
\begin{adjustbox}{center}
    \begin{tabular}{|c|c|c|c|c|c|c|c|c|c|}
        \hline
        Dataset & $c_{333}$ & $c_{321}$ & $c_{322}$ & $c_{331}$ & $c_{332}$ & $c_{341}$ & $c_{342}$ & $c_{343}$ & $c_3$ \\
        \hline\hline
        Science Gallery & 0.15 & 0.00 & 0.03 & 0.00 & 0.02 & 0.00 & 0.01 & 0.19 & 0.00  \\ \hline
        Hospital & 0.28 & 0.00 & 0.01 & 0.00 & 0.00 & 0.00 & 0.00 & 0.16 & 0.00  \\ \hline
        Highschool2012 & 0.35 & 0.00 & 0.01 & 0.00 & 0.01 & 0.00 & 0.00 & 0.24 & 0.00  \\ \hline
        Highschool2013 & 0.34 & 0.00 & 0.01 & 0.00 & 0.01 & 0.00 & 0.00 & 0.13 & 0.00 \\ \hline
        Primaryschool & 0.16 & 0.00 & 0.01 & 0.00 & 0.01 & 0.00 & 0.01 & 0.17 & 0.00 \\ \hline
        Workplace & 0.28 & 0.00 & 0.01 & 0.00 & 0.01 & 0.00 & 0.01 & 0.20 & 0.00 \\ \hline
        Hypertext2009 & 0.35 & 0.00 & 0.02 & 0.00 & 0.01 & 0.00 & 0.00 & 0.05 & 0.00  \\ \hline
        SFHH Conference & 0.37 & 0.00 & 0.01 & 0.00 & 0.01 & 0.00 & 0.00 & 0.15 & 0.00 \\ \hline
    \end{tabular}
    \end{adjustbox}
\caption{Coefficients of the generalized model to predict order $3$ events, when $L=30$ and $\tau=5$. }
\label{tab:coeffs3}
\end{table}

\section{Model evaluation} \label{section-eval}

\subsection{Network prediction quality} \label{section_network_prediction_quality}
Using the optimal coefficients, each model predicts the higher-order network, i.e., events of each order, at any prediction time $t+1$, based on the network observed within $[t-L+1, t]$. The network prediction quality at any prediction time $t+1$ of a given model for events of an arbitrary order is evaluated via the prediction accuracy. This accuracy corresponds to the ratio between the total number of true positives (correctly predicted events) and the total number of events. The prediction quality of a model for events of a given order is the average accuracy over all possible prediction times $t+1 \in [L+1,T]$. \\

\noindent Multiple hyperlinks of a given order $d$ may have the same activation tendency at a given prediction step $t+1$. Given the actual number of order $d$ events at $t+1$, the order $d$ events predicted are, thus, possibly not unique. 
In this case, we consider $100$ realizations of order $d$ events prediction and consider and their average accuracy as the accuracy of order $d$ events at $t+1$. \\

\begin{table}[h]
\begin{adjustbox}{center}
    \begin{tabular}{|c|c c c|c c c|c c c|}
        \hline
        \multirow{2}{*}{Dataset} & \multicolumn{3}{c|}{Order 2} & \multicolumn{3}{c|}{Order 3} & \multicolumn{3}{c|}{Order 4} \\
        \cline{2-10}
         & Baseline & General & Refined & Baseline & General & Refined & Baseline & General & Refined \\
        \hline\hline
        Science Gallery & 0.33 & 0.30 & \hspace{5pt}0.33\hspace{5pt} & 0.15 & \textbf{0.21} & \hspace{5pt}\textbf{0.23}\hspace{5pt} & 0.22 & \textbf{0.59} & \hspace{5pt}\textbf{0.58}\hspace{5pt} \\ \hline
        Hospital & 0.52 & 0.52 & \hspace{5pt}\textbf{0.54}\hspace{5pt} & 0.32 & \textbf{0.49} & \hspace{5pt}\textbf{0.50}\hspace{5pt} & 0.17 & \textbf{0.74} & \hspace{5pt}\textbf{0.72}\hspace{5pt} \\ \hline
        Highschool2012 & 0.55 & 0.52 & \hspace{5pt}\textbf{0.56}\hspace{5pt} & 0.38 & \textbf{0.47} & \hspace{5pt}\textbf{0.49}\hspace{5pt} & 0.19 & \textbf{0.60} & \hspace{5pt}\textbf{0.61}\hspace{5pt} \\ \hline
        Highschool2013 & 0.61 & 0.60 & \hspace{5pt}0.61\hspace{5pt} & 0.35 & \textbf{0.39} & \hspace{5pt}\textbf{0.40}\hspace{5pt} & 0.36 & \textbf{0.59} & \hspace{5pt}\textbf{0.61}\hspace{5pt} \\ \hline
        Primaryschool & 0.32 & 0.31 & \hspace{5pt}0.32\hspace{5pt} & 0.16 & \textbf{0.17} & \hspace{5pt}\textbf{0.19}\hspace{5pt} & 0.08 & \textbf{0.32} & \hspace{5pt}\textbf{0.30}\hspace{5pt} \\ \hline
        Workplace & 0.56 & 0.55 & \hspace{5pt}\textbf{0.57}\hspace{5pt} & 0.31 & \textbf{0.47} & \hspace{5pt}\textbf{0.49}\hspace{5pt} & 0.07 & \textbf{0.50} & \hspace{5pt}\textbf{0.43}\hspace{5pt} \\ \hline
        Hypertext2009 & 0.50 & 0.49 & \hspace{5pt}0.50\hspace{5pt} & 0.34 & \textbf{0.51} & \hspace{5pt}\textbf{0.52}\hspace{5pt} & 0.12 & \textbf{0.65} & \hspace{5pt}\textbf{0.59}\hspace{5pt} \\ \hline
        SFHH Conference & 0.52 & 0.52 & \hspace{5pt}\textbf{0.53}\hspace{5pt} & 0.38 & \textbf{0.43} & \hspace{5pt}\textbf{0.45}\hspace{5pt} & 0.40 & \textbf{0.57} & \hspace{5pt}\textbf{0.57}\hspace{5pt} \\ \hline
    \end{tabular}
    \end{adjustbox}
\caption{Prediction accuracy of all three models, with $L=30$ and $\tau=5$, for $2$-, $3$-, and $4$-hyperlinks in every dataset. The prediction quality of the generalized model and/or refined model is displayed in bold if it exceeds the baseline's performance.}
\label{tab:Pred_results_combined0}
\end{table}

\noindent The prediction quality of the baseline model, generalized model and refined model, produced with $L=30$ and $\tau=5$, is displayed in Table \ref{tab:Pred_results_combined0}. We find that the generalized model and refined model perform in general better than the baseline. This out-performance is more evident when predicting order $3$ and order $4$ events.
The refined model performs better than the generalized model when predicting order $2$ and order $3$ events. This supports that the refined model captures the contribution of key types of neighboring hyperlinks when predicting the activity of an arbitrary target link. However, the refined model and generalized model perform similarly when predicting order $4$ events. This is because the number of order $4$ events is small and the design of the refined model is driven by the observation that the sub- and super- hyperlinks contribute more, reflected in their corresponding coefficients in the generalized model, in predicting order $2$ or $3$ events than other types of neighboring links, which is less evident when predicting order $4$ events.
The less evident out-performance of the refined model than the baseline in predicting order $2$ events is because most order $2$ target links have no super- nor sub- neighboring hyperlink.

\subsection{Performance analysis of the refined model} \label{section_performance_analysis_refined}
\subsubsection{Influence of coefficients $c_\phi$}
\begin{figure*}[h!]
    \centering
    \begin{subfigure}[b]{0.49\textwidth}
        \includegraphics[width=\textwidth]{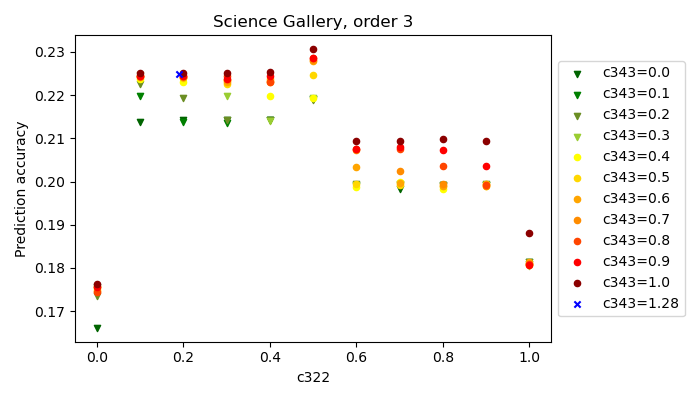}
        \label{fig:SG_c322}
    \end{subfigure}
    \begin{subfigure}[b]{0.49\textwidth}
        \includegraphics[width=\textwidth]{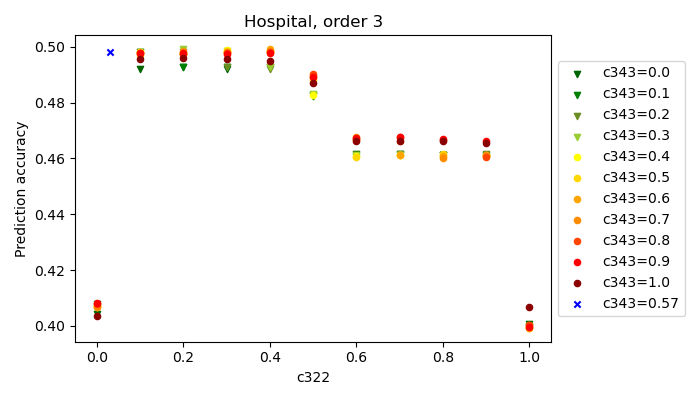}
        \label{fig:HOSP_c322}
    \end{subfigure}
    \begin{subfigure}[b]{0.49\textwidth}
        \includegraphics[width=\textwidth]{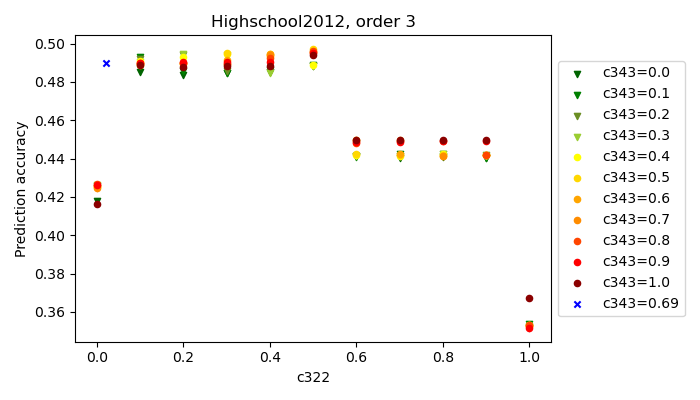}
        \label{fig:HS2012_c322}
    \end{subfigure}
    \begin{subfigure}[b]{0.49\textwidth}
        \includegraphics[width=\textwidth]{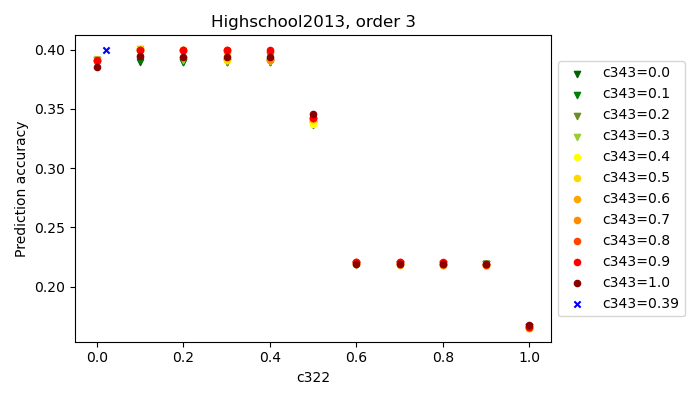}
        \label{fig:HS2013_c322}
    \end{subfigure}
    \begin{subfigure}[b]{0.49\textwidth}
        \includegraphics[width=\textwidth]{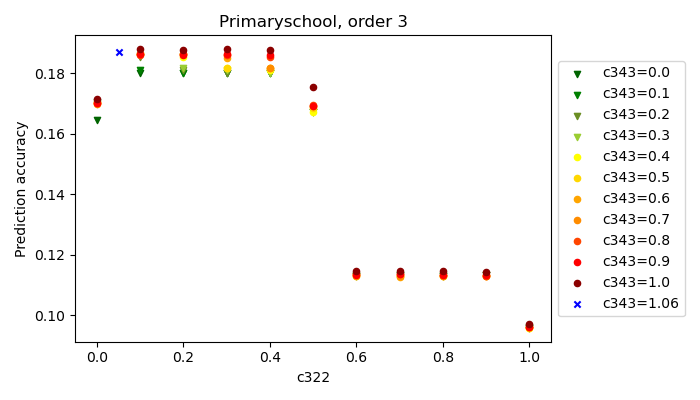}
        \label{fig:PS_c322}
    \end{subfigure}
    \begin{subfigure}[b]{0.49\textwidth}
        \includegraphics[width=\textwidth]{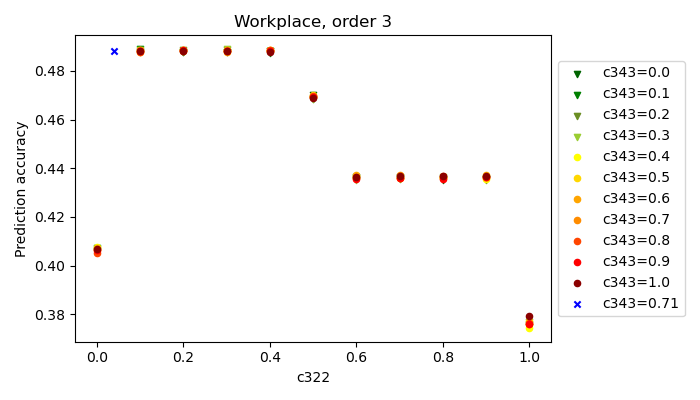}
        \label{fig:WORK_c322}
    \end{subfigure}
    \begin{subfigure}[b]{0.49\textwidth}
        \includegraphics[width=\textwidth]{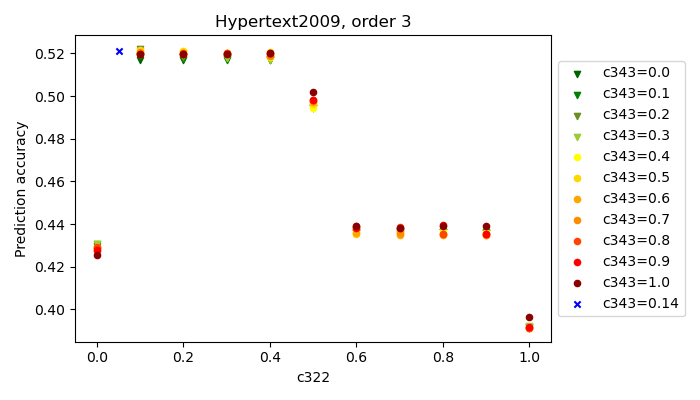}
        \label{fig:HT09_c322}
    \end{subfigure}
    \begin{subfigure}[b]{0.49\textwidth}
        \includegraphics[width=\textwidth]{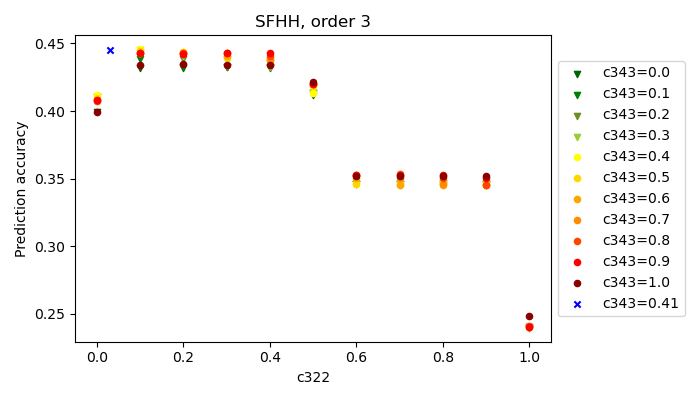}
        \label{fig:SFHH_c322}
    \end{subfigure}
    \caption{Prediction accuracy of order $3$ events in the refined model as a function of $c_{322}$, with $c_{343}$ fixed, for all datasets.}
    \label{fig:coeffs_range_3}
\end{figure*}
We previously computed the prediction accuracy for our models by using an optimal set of coefficients, learned via Lasso regression, based on the temporal network observed over the complete time span $[1,T]$. In reality, the network at the prediction step and further in time, thus within $[t+1,T]$, is unknown. Hence, we aim to explore the influence of the coefficients $c_\phi$ on the performance of our refined model. This allows us to explore which coefficient ranges tend to lead to close to optimal prediction. \\

\noindent We calculate the prediction accuracy of our refined model, where we fix $c_{ddd}=1$, and the coefficients associated with the sub- and super-hyperlink with all possible values within $[0.0, 0.1, ..., 1.0] \times [0.0, 0.1, ..., 1.0]$. We take the prediction of order $3$ events as an example. Figure \ref{fig:coeffs_range_3} shows the prediction accuracy as a function of the sub- and super-coefficients. We find that that close to optimal prediction accuracy is obtained approximately when $c_{322} \in [ 0.1, 0.4 ]<c_{333}=1$. 
This is in line with our finding in the generalized model in  Section \ref{section_network_prediction_quality} that the past activity of a $\phi=(322)$-neighbor contributes less to the prediction of the future activity of an order $3$ target link than the past activity of the target itself. 
The influence of the coefficient $c_{343}$ for super-hyperlinks on the prediction quality is in general less than that of the coefficient $c_{322}$ for sub-hyperlinks. In a few networks like Workplace and Hypertext2009, the prediction accuracy does not change with the choice of $c_{343}$. In general, a $c_{343} \in [0.8, 1]$ tends to enable close to optimal prediction accuracy. \\

\noindent We have also added the coefficients of the refined model learned from Lasso Regression to Figure \ref{fig:coeffs_range_3}. The parameter analysis in Figure \ref{fig:coeffs_range_3} assumes $c_{333}=1$. Hence, coefficients $c_{322}$ and $c_{343}$ learned from Lasso normalized by the learned $c_{333}$, respectively, together with their corresponding prediction accuracy are included Figure \ref{fig:coeffs_range_3}. The refined model with the learned coefficients seems to lead to the optimal prediction quality in all networks, but close to optimal in Science Gallery. This is because Lasso Regression aims not solely to optimize the prediction accuracy but also to minimize the magnitude of coefficients. For the learned coefficient, we observe that $c_{322}<c_{343}$. Consistently, we found the ranges $c_{322} \in [ 0.1, 0.4 ]$ and $c_{343} \in [0.8, 1]$ that tend to lead to close optimal prediction, which also indicates $c_{322}<c_{343}$. \\

\begin{figure*}[h!]
    \centering
    \begin{subfigure}[b]{0.49\textwidth}
        \includegraphics[width=\textwidth]{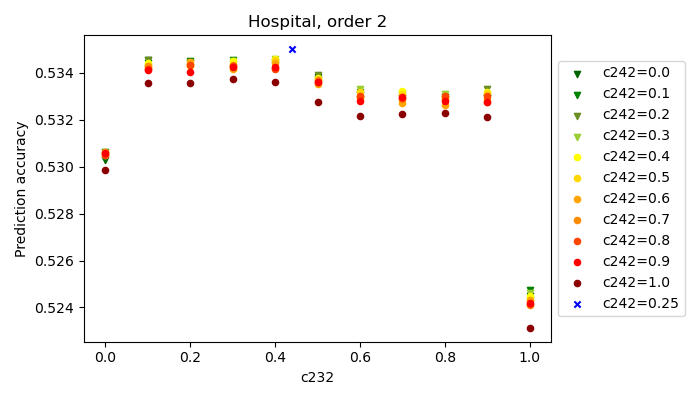}
        \label{fig:HOSP_c232}
    \end{subfigure}
    \begin{subfigure}[b]{0.49\textwidth}
        \includegraphics[width=\textwidth]{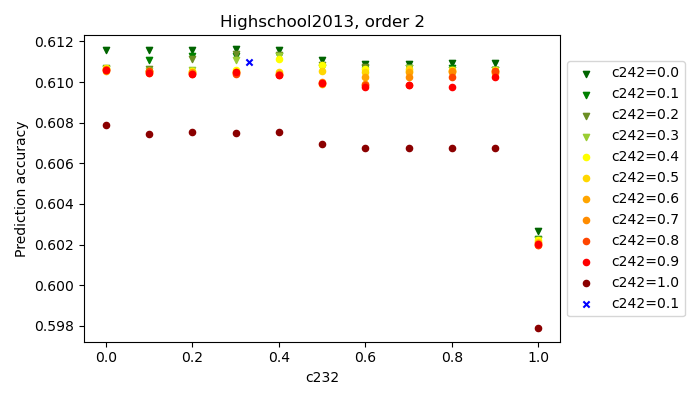}
        \label{fig:SFHH_c232}
    \end{subfigure}
    \begin{subfigure}[b]{0.49\textwidth}
        \includegraphics[width=\textwidth]{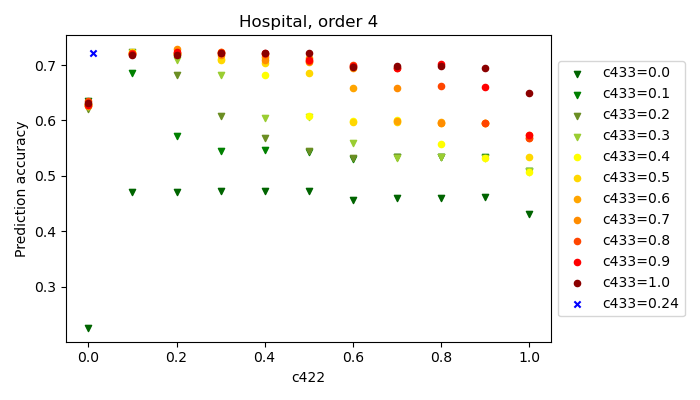}
        \label{fig:HOSP_c422}
    \end{subfigure}
    \begin{subfigure}[b]{0.49\textwidth}
        \includegraphics[width=\textwidth]{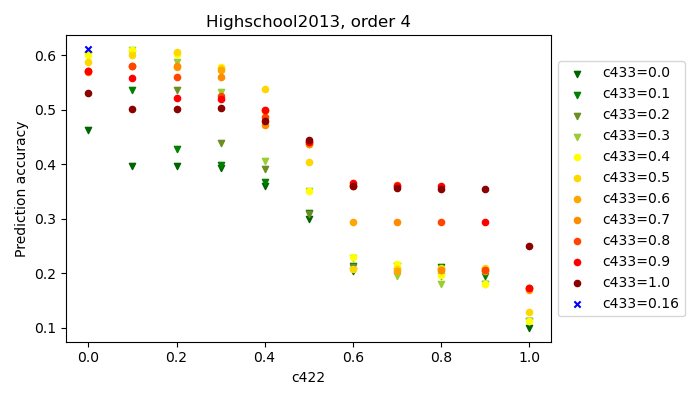}
        \label{fig:SFHH_c422}
    \end{subfigure}
    \caption{Prediction accuracy for events of order $2$ or $4$ in the refined model as a function of one coefficient, with the other coefficient fixed, for two datasets.}
    \label{fig:coeffs_range_2_and_4}
\end{figure*}
\noindent For events of order $2$, we find that the prediction quality tends to be optimal when $c_{232},c_{242}\in[0.1, 0.4]$. However, their influence on the prediction quality is less evident, as partially shown in Figure \ref{fig:coeffs_range_2_and_4}. For order $4$ events (see also Figure \ref{fig:coeffs_range_2_and_4}), $c_{422}$ and $c_{433}$ affect the prediction quality evidently. We find that $c_{422} \in [ 0.1, 0.4 ]$ and $c_{433} \in [0.6, 0.9]$ tend to give rise to close to optimal prediction precision. Hence, to achieve the optimal prediction quality, we could choose $c_{422}<c_{433}<c_{444}$ and $c_{322}<c_{343} \leq c_{333}$, thus let a hyperlink that overlaps more with the target hyperlink in nodes to contribute more. This result is complementary to our finding in Section \ref{section_phi_memory}, that among the super- and sub-hyperlinks, those that overlap more with a target hyperlink in nodes are more correlated with the target.
\subsubsection{Influence of $\tau$ and $L$} \label{model-parameters}
\begin{figure*}[h!]
    \centering
    \begin{subfigure}[b]{0.45\textwidth}
        \includegraphics[width=\textwidth]{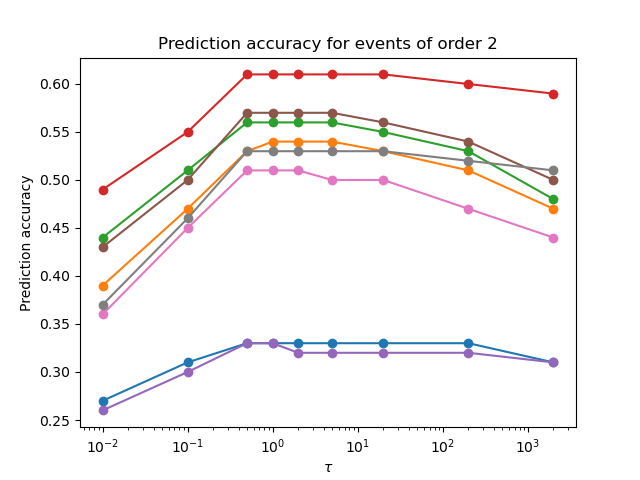}
        \caption{$d = 2$}
        \label{fig:tau_acc_order2}
    \end{subfigure}
    \begin{subfigure}[b]{0.45\textwidth}
        \includegraphics[width=\textwidth]{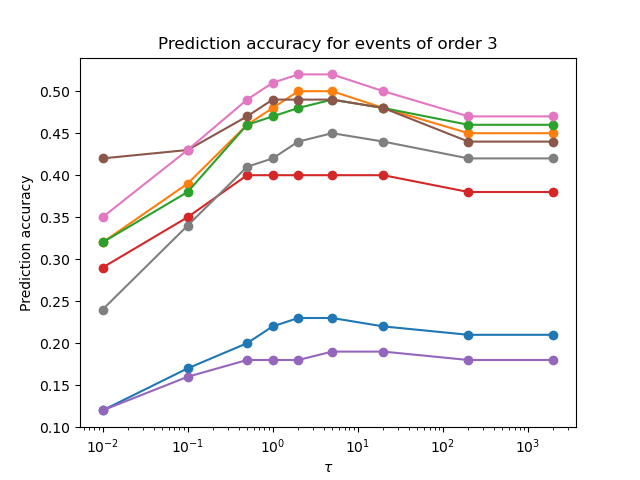}
        \caption{$d = 3$}
        \label{fig:tau_acc_order3}
    \end{subfigure}
    \begin{subfigure}[b]{0.45\textwidth}
        \includegraphics[width=\textwidth]{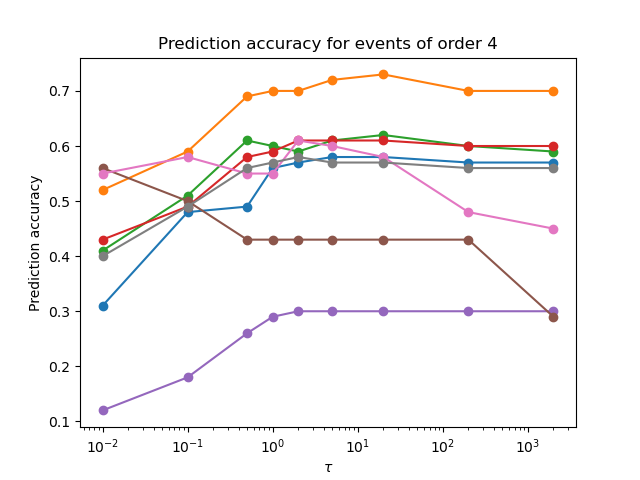}
        \caption{$d = 4$}
        \label{fig:tau_acc_order4}
    \end{subfigure}
    \begin{subfigure}[b]{0.45\textwidth}
        \includegraphics[width=\textwidth]{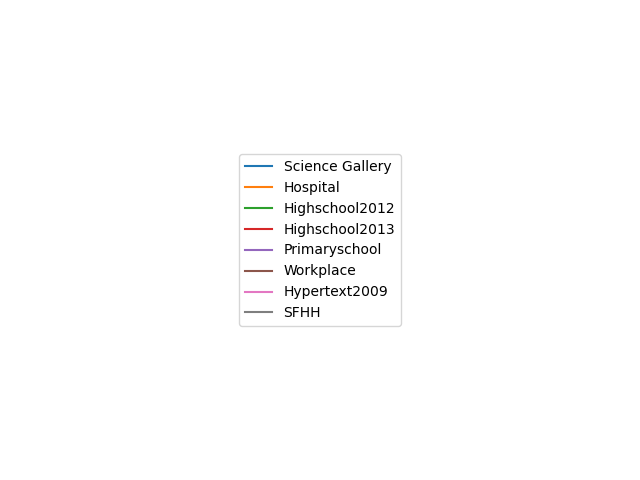}
    \end{subfigure}
    \caption{Prediction accuracy of hyperlinks of order $d \in [2,4]$ as a function of the exponential decay factor $\tau$ in eight real-world physical contact networks.}
    \label{fig:tau_acc}
\end{figure*}
It has been shown that $\tau\in[0.5,5]$ leads to an optimal prediction accuracy for pairwise temporal network prediction \cite{zou2023memory}. Thus there is no need to calibrate $\tau$. In higher-order temporal, would a broad range of $\tau$ all lead to optimal accuracy? 
We now compare the prediction accuracy of the refined model, for distinct values of $\tau\in[10^{-2},10^3]$ when $L=30$. Given $L=30$ and a given $\tau$, the coefficients of the refined model are fitted via Lasso Regression and further used to predict higher-order events. Figure \ref{fig:tau_acc} shows that approximately, any $\tau\in[0.5,5]$ leads to close to optimal prediction quality in all networks considered, meaning the parameter $\tau$ could be chosen arbitrarily within this broad range without the need to be calibrated. Secondly, we generally find a less optimal performance for large values of $\tau$. This is because a large $\tau$ results in a fast decay of the exponential in Equation (\ref{eq:xphi}), thereby suppressing the contribution of previous events. Lastly, we find that the worst performance is achieved for small values of $\tau$, because when $\tau$ is small, previous interactions almost contribute equally in event prediction, independent of when these interactions occur. \\
\begin{figure*}[h!]
    \centering
    \begin{subfigure}[b]{0.45\textwidth}
        \includegraphics[width=\textwidth]{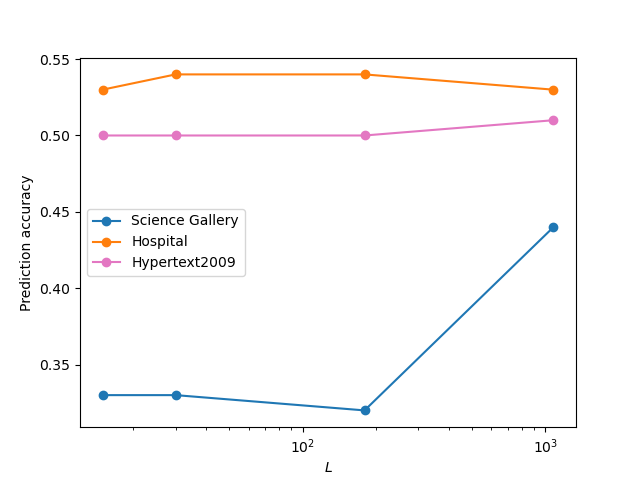}
        \caption{$d = 2$}
        \label{fig:L_acc_order2}
    \end{subfigure}
    \begin{subfigure}[b]{0.45\textwidth}
        \includegraphics[width=\textwidth]{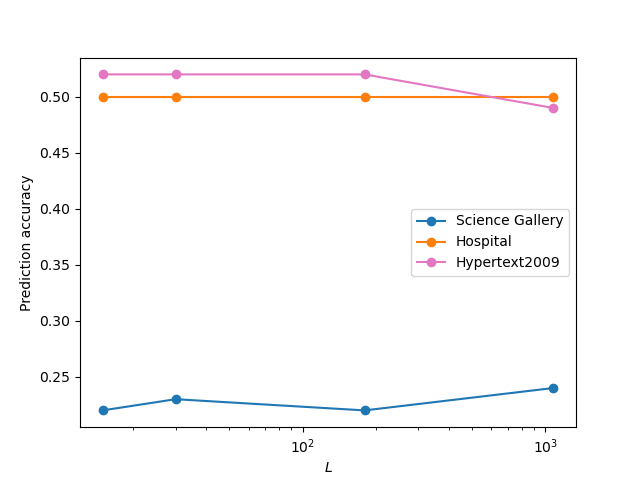}
        \caption{$d = 3$}
        \label{fig:L_acc_order3}
    \end{subfigure}
    \begin{subfigure}[b]{0.45\textwidth}
        \includegraphics[width=\textwidth]{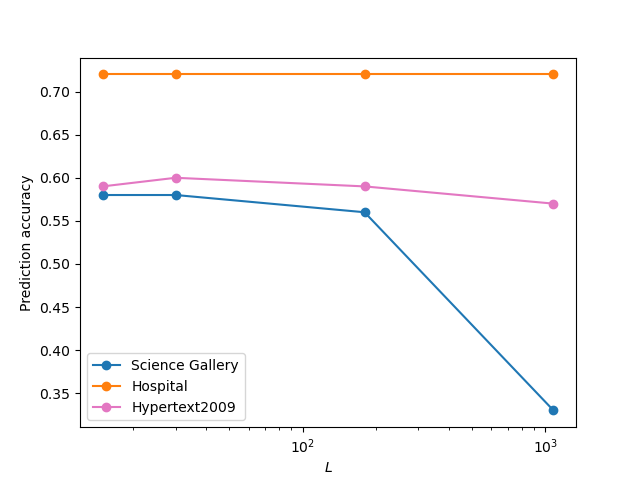}
        \caption{$d = 4$}
        \label{fig:L_acc_order4}
    \end{subfigure}
    \caption{Prediction accuracy of hyperlinks of order $d \in [2,4]$ as a function of the prediction time span $L$ in three real-world physical contact networks.}
    \label{fig:L_acc}
\end{figure*}

\noindent As mentioned earlier, the choice of $L$ could be driven by the context of real-world prediction problem, e.g., how long of the network in the past is available and can be used for the prediction. A small $L$ like $L=30$ is more feasible in practice and leads to a lower computational complexity. If a temporal network can be observed for long in the past for the prediction of future events, would the choice of $L$ influence the prediction quality? 
In Figure \ref{fig:L_acc} we explore the prediction accuracy of events of order $d\in[2,4]$ as a function of $L\in[15,30,180,1080]$ while keeping a fixed $\tau=5$. A $L=30$ means an observation of a network for $600s$ in the past is used for the prediction, as the duration of each time step is $20s$. Hence, the set of $L$ considered implies an observation of a duration ranging from $5$ minutes to $6$ hours to predict the network $20s$ ahead. Figure \ref{fig:L_acc} shows that, in three network example, the prediction quality is relatively stable as $L$ varies within $L\in[15,30,180]$. When $L$ is increased to the extreme $L=1080$, the prediction accuracy could be higher or lower. Hence, $L\in[15,30,180,1080]$ seems to be a good choice, for which data is relatively accessible and the computational complexity is low.

\section{Conclusion and Discussion}
In this work, we propose network-based higher-order temporal network prediction models that estimate the activity of each target hyperlink at the next time step, based on the past activity of this target and of its $\phi$-neighboring hyperlinks. The generalized model considers the contribution of all types $\phi$ of neighboring hyperlink, motivated by the memory in activity of each hyperlink and between neighboring hyperlinks. The refined model considers only two types of neighboring links, the super- and sub-links, inspired by the coefficients analysis in the generalized model and the observation that a hyperlink is more strongly correlated with these two types of neighbors than others. \\

\noindent We discover that the generalized and refined models consistently outperform the baseline derived from a pairwise prediction method. The refined model performs better than the generalized model in the prediction of $2$- and $3$-events. The optimal coefficients of our models learned via Lasso Regression as well as the correlation analysis between neighboring hyperlinks in activity unravel that the past activity of the target link itself contributes the most in the prediction, followed by the activity of super- and sub-links. Among the super- and sub-links, those overlap more with the target link in node contribute more in the prediction. \\

\noindent The identification of the optimal coefficients for the refined model is via the learning from the temporal network in the past as well as in the future, which is unrealistic. Hence, we analyze the influence of various parameters on the prediction accuracy, and identify ranges of parameters that may lead to close to optimal prediction quality. An interesting future work is to explore whether the model thus the optimal coefficients learned from the network observed in the past, could well predict future events.\\

\noindent In this work, we have focused on higher-order social contact networks that are derived from measurements of pairwise interactions. These networks have the property that a hyperlink and its sub-links are never activated at the same time. It is interesting to explore methods to predict other types of higher-order temporal networks that may not have this property nor the memory property, e.g. collaboration networks \cite{abella2023unraveling}. Moreover, better prediction accuracy can possibly be achieved when a different decay factor $\tau$ is considered for each type $\phi$ of neighbors, because the slope of the time-decaying memory between a target link and a neighboring link depends on the type of the neigbhoring link. Finally, the baseline model could be improved by, e.g., using the total number of events of each order in the prediction step, which has already been utilized by our models. \\

\vspace{-2mm}
\section*{Acknowlegement}
We thank for the support of Netherlands Organisation for Scientific Research NWO (TOP Grant no. 612.001.802, project FORT-PORT no. KICH1.VE03.21.008) and NExTWORKx, a collaboration between TU Delft and KPN on future telecommunication networks.
\bibliographystyle{unsrt}
\bibliography{bibliography}

\begin{thebibliography}{10}

\bibitem{holme2012temporal}
Petter Holme and Jari Saram{\"a}ki.
\newblock Temporal networks.
\newblock {\em Physics reports}, 519(3):97--125, 2012.

\bibitem{masuda2016guide}
Naoki Masuda and Renaud Lambiotte.
\newblock {\em A guide to temporal networks}.
\newblock World Scientific, 2016.

\bibitem{holme2015modern}
Petter Holme.
\newblock Modern temporal network theory: a colloquium.
\newblock {\em The European Physical Journal B}, 88:1--30, 2015.

\bibitem{battiston2020networks}
Federico Battiston, Giulia Cencetti, Iacopo Iacopini, Vito Latora, Maxime Lucas, Alice Patania, Jean-Gabriel Young, and Giovanni Petri.
\newblock Networks beyond pairwise interactions: structure and dynamics.
\newblock {\em Physics Reports}, 874:1--92, 2020.

\bibitem{battiston2021physics}
Federico Battiston et~al.
\newblock The physics of higher-order interactions in complex systems.
\newblock {\em Nature Physics}, 17(10):1093--1098, 2021.

\bibitem{sekara2016fundamental}
Vedran Sekara, Arkadiusz Stopczynski, and Sune Lehmann.
\newblock Fundamental structures of dynamic social networks.
\newblock {\em Proceedings of the national academy of sciences}, 113(36):9977--9982, 2016.

\bibitem{patania2017shape}
Alice Patania, Giovanni Petri, and Francesco Vaccarino.
\newblock The shape of collaborations.
\newblock {\em EPJ Data Science}, 6:1--16, 2017.

\bibitem{lu2012recommender}
Linyuan L{\"u}, Mat{\'u}{\v{s}} Medo, Chi~Ho Yeung, Yi-Cheng Zhang, Zi-Ke Zhang, and Tao Zhou.
\newblock Recommender systems.
\newblock {\em Physics reports}, 519(1):1--49, 2012.

\bibitem{aleta2020link}
Alberto Aleta, Marta Tuninetti, Daniela Paolotti, Yamir Moreno, and Michele Starnini.
\newblock Link prediction in multiplex networks via triadic closure.
\newblock {\em Physical Review Research}, 2(4):042029, 2020.

\bibitem{zhou2019dynamic}
Yujing Zhou, Yang Pei, Yuanye He, Jingjie Mo, Jiong Wang, and Neng Gao.
\newblock Dynamic graph link prediction by semantic evolution.
\newblock In {\em ICC 2019-2019 IEEE International Conference on Communications (ICC)}, pages 1--6. IEEE, 2019.

\bibitem{li2014deep}
Xiaoyi Li, Nan Du, Hui Li, Kang Li, Jing Gao, and Aidong Zhang.
\newblock A deep learning approach to link prediction in dynamic networks.
\newblock In {\em Proceedings of the 2014 SIAM International conference on data mining}, pages 289--297. SIAM, 2014.

\bibitem{chen2019generative}
Jinyin Chen, Xiang Lin, Chenyu Jia, Yuwei Li, Yangyang Wu, Haibin Zheng, and Yi~Liu.
\newblock Generative dynamic link prediction.
\newblock {\em Chaos: An Interdisciplinary Journal of Nonlinear Science}, 29(12):123111, 2019.

\bibitem{chen2019lstm}
Jinyin Chen, Jian Zhang, Xuanheng Xu, Chenbo Fu, Dan Zhang, Qingpeng Zhang, and Qi~Xuan.
\newblock {E-LSTM-D}: {A} deep learning framework for dynamic network link prediction.
\newblock {\em IEEE Transactions on Systems, Man, and Cybernetics: Systems}, 51(6):3699--3712, 2019.

\bibitem{zou2023memory}
Li~Zou, An~Wang, and Huijuan Wang.
\newblock Memory based temporal network prediction.
\newblock In {\em Complex Networks and Their Applications XI: Proceedings of The Eleventh International Conference on Complex Networks and their Applications: COMPLEX NETWORKS 2022—Volume 2}, pages 661--673. Springer, 2023.

\bibitem{benson2018simplicial}
Austin~R Benson, Rediet Abebe, Michael~T Schaub, Ali Jadbabaie, and Jon Kleinberg.
\newblock Simplicial closure and higher-order link prediction.
\newblock {\em Proceedings of the National Academy of Sciences}, 115(48):E11221--E11230, 2018.

\bibitem{liu2023higher}
Bo~Liu, Rongmei Yang, and Linyuan L{\"u}.
\newblock Higher-order link prediction via local information.
\newblock {\em Chaos 1 August 2023; 33 (8): 083108}, 2023.

\bibitem{piaggesi2022effective}
Simone Piaggesi, Andr{\'e} Panisson, and Giovanni Petri.
\newblock Effective higher-order link prediction and reconstruction from simplicial complex embeddings.
\newblock In {\em Learning on Graphs Conference}, pages 55--1. PMLR, 2022.

\bibitem{liu2022neural}
Yunyu Liu, Jianzhu Ma, and Pan Li.
\newblock Neural predicting higher-order patterns in temporal networks.
\newblock In {\em Proceedings of the ACM Web Conference 2022}, pages 1340--1351, 2022.

\bibitem{cencetti2021temporal}
Giulia Cencetti, Federico Battiston, Bruno Lepri, and M{\'a}rton Karsai.
\newblock Temporal properties of higher-order interactions in social networks.
\newblock {\em Scientific reports}, 11(1):7028, 2021.

\bibitem{gallo2023higher}
Luca Gallo, Lucas Lacasa, Vito Latora, and Federico Battiston.
\newblock Higher-order correlations reveal complex memory in temporal hypergraphs.
\newblock {\em Nature Communications}, 15, 06 2024.

\bibitem{ceria2023temporal}
Alberto Ceria and Huijuan Wang.
\newblock Temporal-topological properties of higher-order evolving networks.
\newblock {\em Scientific Reports}, 13(1):5885, 2023.

\bibitem{iacopini2023temporal}
Iacopo Iacopini, M{\'a}rton Karsai, and Alain Barrat.
\newblock The temporal dynamics of group interactions in higher-order social networks.
\newblock {\em arXiv preprint arXiv:2306.09967}, 2023.

\bibitem{jung-muller2024higher}
Mathieu Jung-Muller, Alberto Ceria, and Huijuan Wang.
\newblock Higher-order temporal network prediction.
\newblock In Hocine Cherifi, Luis~M. Rocha, Chantal Cherifi, and Murat Donduran, editors, {\em Complex Networks {\&} Their Applications XII}, pages 461--472, Cham, 2024. Springer Nature Switzerland.

\bibitem{fournet2014contact}
Julie Fournet and Alain Barrat.
\newblock Contact patterns among high school students.
\newblock {\em PloS one}, 9(9):e107878, 2014.

\bibitem{mastrandrea2015contact}
Rossana Mastrandrea, Julie Fournet, and Alain Barrat.
\newblock Contact patterns in a high school: a comparison between data collected using wearable sensors, contact diaries and friendship surveys.
\newblock {\em PloS one}, 10(9):e0136497, 2015.

\bibitem{stehle2011high}
Juliette Stehl{\'e}, Nicolas Voirin, Alain Barrat, Ciro Cattuto, Lorenzo Isella, Jean-Fran{\c{c}}ois Pinton, Marco Quaggiotto, Wouter Van~den Broeck, Corinne R{\'e}gis, Bruno Lina, et~al.
\newblock High-resolution measurements of face-to-face contact patterns in a primary school.
\newblock {\em PloS one}, 6(8):e23176, 2011.

\bibitem{genois2018can}
Mathieu G{\'e}nois and Alain Barrat.
\newblock Can co-location be used as a proxy for face-to-face contacts?
\newblock {\em EPJ Data Science}, 7(1):1--18, 2018.

\bibitem{isella2011s}
Lorenzo Isella, Juliette Stehl{\'e}, Alain Barrat, Ciro Cattuto, Jean-Fran{\c{c}}ois Pinton, and Wouter Van~den Broeck.
\newblock What's in a crowd? analysis of face-to-face behavioral networks.
\newblock {\em Journal of theoretical biology}, 271(1):166--180, 2011.

\bibitem{vanhems2013estimating}
Philippe Vanhems, Alain Barrat, Ciro Cattuto, Jean-Fran{\c{c}}ois Pinton, Nagham Khanafer, Corinne R{\'e}gis, Byeul-a Kim, Brigitte Comte, and Nicolas Voirin.
\newblock Estimating potential infection transmission routes in hospital wards using wearable proximity sensors.
\newblock {\em PloS one}, 8(9):e73970, 2013.

\bibitem{ceria2022topological}
Alberto Ceria, Shlomo Havlin, Alan Hanjalic, and Huijuan Wang.
\newblock Topological--temporal properties of evolving networks.
\newblock {\em Journal of Complex Networks}, 10(5):cnac041, 2022.

\bibitem{abella2023unraveling}
David Abella, Piero Birello, Leonardo Di~Gaetano, Sara Ghivarello, Narayan~G Sabhahit, Christel Sirocchi, and Juan Fern{\'a}ndez-Gracia.
\newblock Unraveling higher-order dynamics in collaboration networks.
\newblock {\em arXiv preprint arXiv:2306.17521}, 2023.

\end{thebibliography}

\section*{Appendix} \label{section-appendix}
\refstepcounter{section}
\counterwithin{figure}{section}

\renewcommand{\thefigure}{A\arabic{figure}}
\setcounter{figure}{0}

\begin{figure*}[h!]
    \centering
    \begin{subfigure}[b]{0.4\textwidth}
        \includegraphics[width=\textwidth]{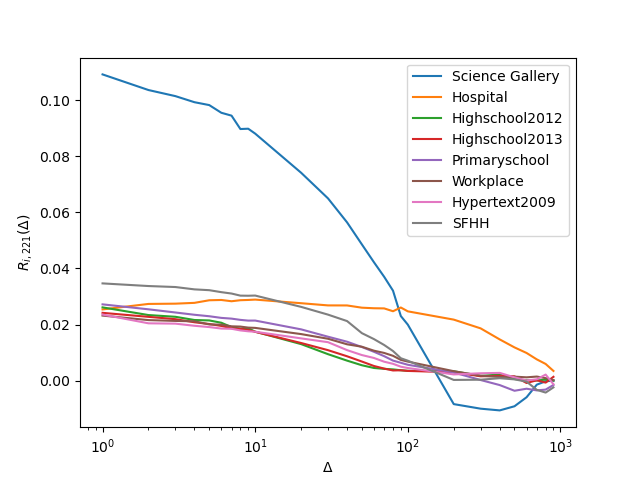}
        \caption{$\phi = 221$}
        \label{fig:pearsoncor_221}
    \end{subfigure}
    \begin{subfigure}[b]{0.4\textwidth}
        \includegraphics[width=\textwidth]{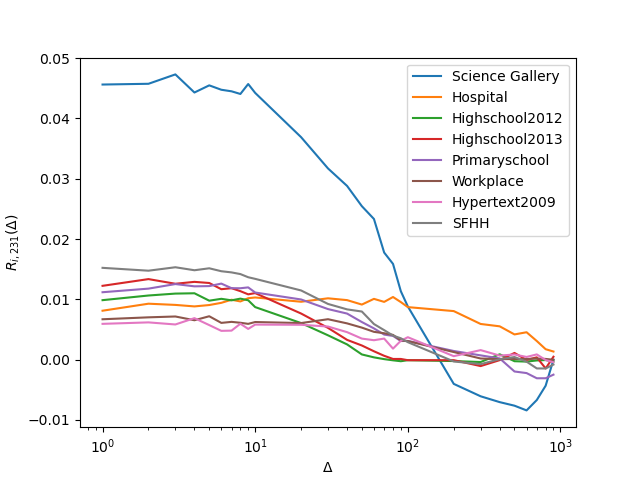}
        \caption{$\phi = 231$}
        \label{fig:pearsoncor_231}
    \end{subfigure}
    \begin{subfigure}[b]{0.4\textwidth}
        \includegraphics[width=\textwidth]{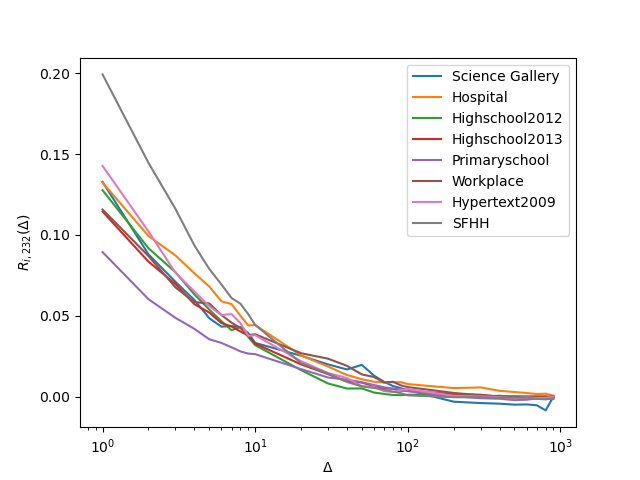}
        \caption{$\phi = 232$}
        \label{fig:pearsoncor_232}
    \end{subfigure}
    \begin{subfigure}[b]{0.4\textwidth}
        \includegraphics[width=\textwidth]{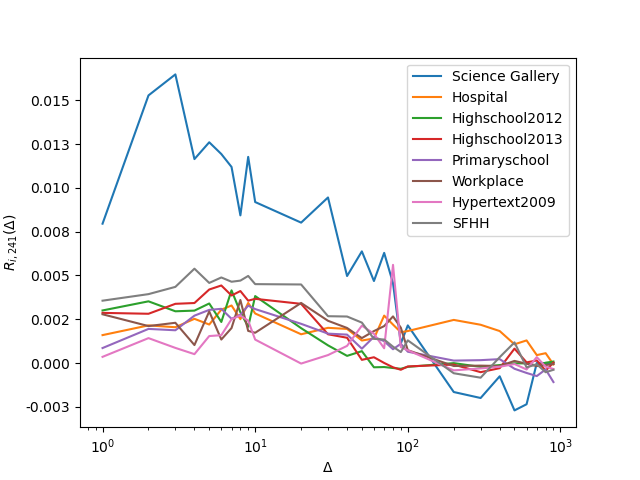}
        \caption{$\phi = 241$}
        \label{fig:pearsoncor_241}
    \end{subfigure}
    \begin{subfigure}[b]{0.4\textwidth}
        \includegraphics[width=\textwidth]{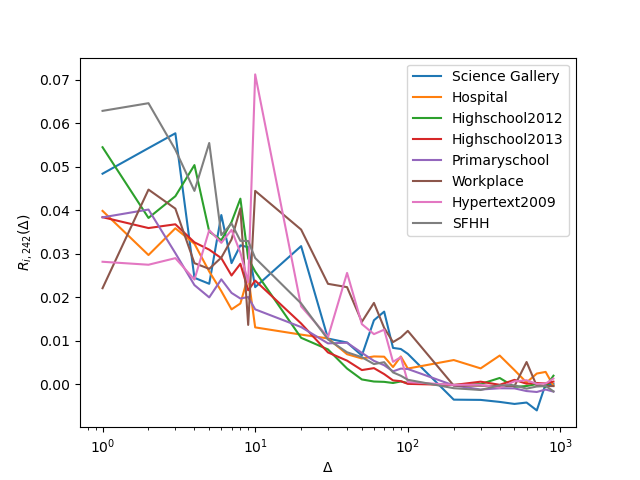}
        \caption{$\phi = 242$}
        \label{fig:pearsoncor_242}
    \end{subfigure}
    \caption{Average Pearson correlation coefficient $R_{i,\phi}(\Delta)$ for order $2$ hyperlinks in eight real-world physical contact networks as a function of time lag $\Delta$.}
    \label{fig:pearsoncor_rest2}
\end{figure*}

\begin{figure*}[h!]
    \centering
    \begin{subfigure}[b]{0.4\textwidth}
        \includegraphics[width=\textwidth]{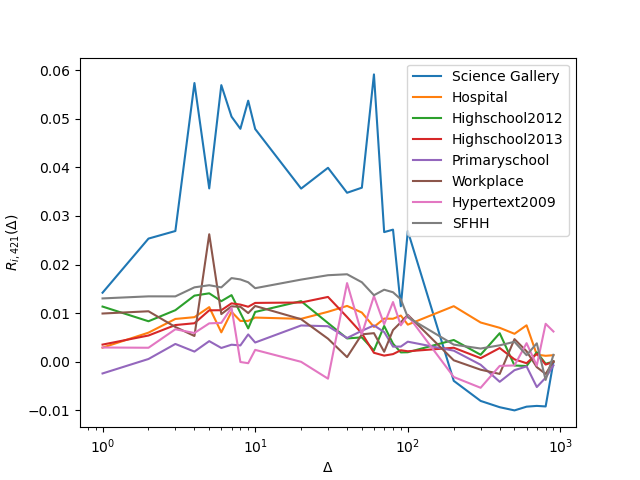}
        \caption{$\phi = 421$}
        \label{fig:pearsoncor_421}
    \end{subfigure}
    \begin{subfigure}[b]{0.4\textwidth}
        \includegraphics[width=\textwidth]{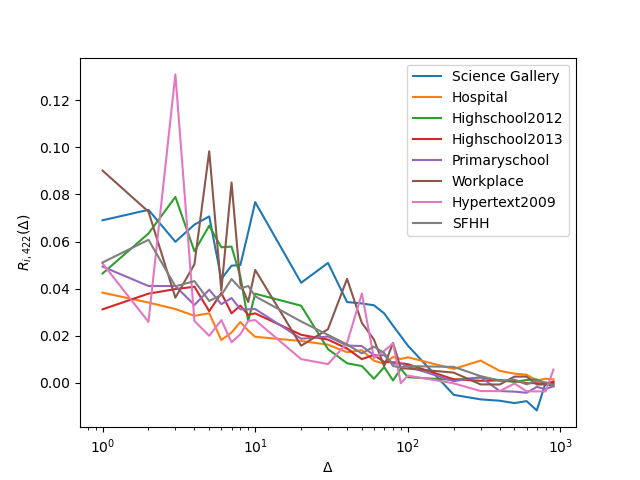}
        \caption{$\phi = 422$}
        \label{fig:pearsoncor_422}
    \end{subfigure}
    \begin{subfigure}[b]{0.4\textwidth}
        \includegraphics[width=\textwidth]{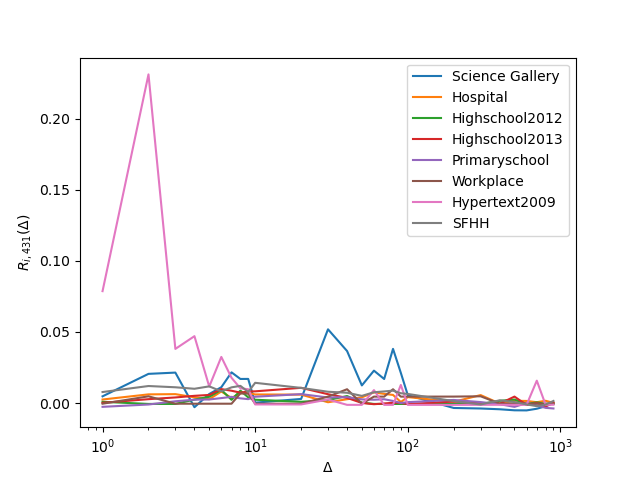}
        \caption{$\phi = 431$}
        \label{fig:pearsoncor_431}
    \end{subfigure}
    \begin{subfigure}[b]{0.4\textwidth}
        \includegraphics[width=\textwidth]{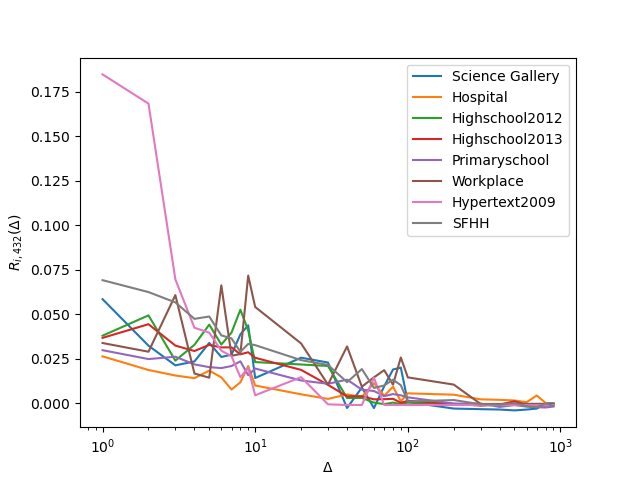}
        \caption{$\phi = 432$}
        \label{fig:pearsoncor_432}
    \end{subfigure}
    \begin{subfigure}[b]{0.4\textwidth}
        \includegraphics[width=\textwidth]{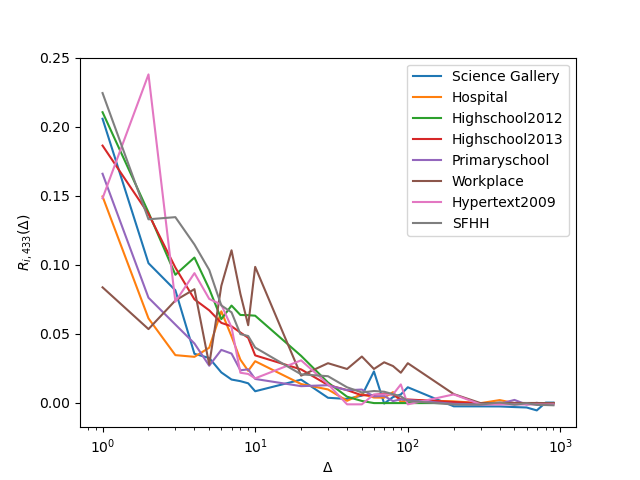}
        \caption{$\phi = 433$}
        \label{fig:pearsoncor_433}
    \end{subfigure}
    \begin{subfigure}[b]{0.4\textwidth}
        \includegraphics[width=\textwidth]{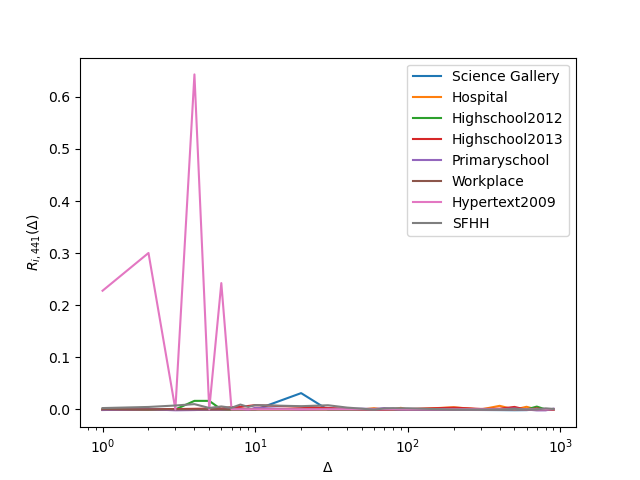}
        \caption{$\phi = 441$}
        \label{fig:pearsoncor_441}
    \end{subfigure}
    \begin{subfigure}[b]{0.4\textwidth}
        \includegraphics[width=\textwidth]{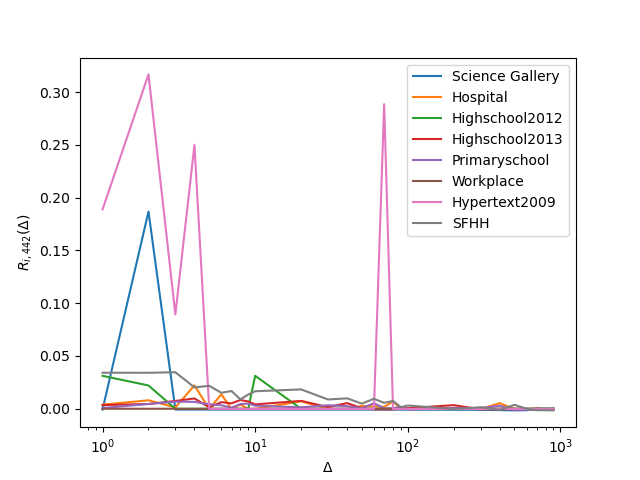}
        \caption{$\phi = 442$}
        \label{fig:pearsoncor_442}
    \end{subfigure}
    \begin{subfigure}[b]{0.4\textwidth}
        \includegraphics[width=\textwidth]{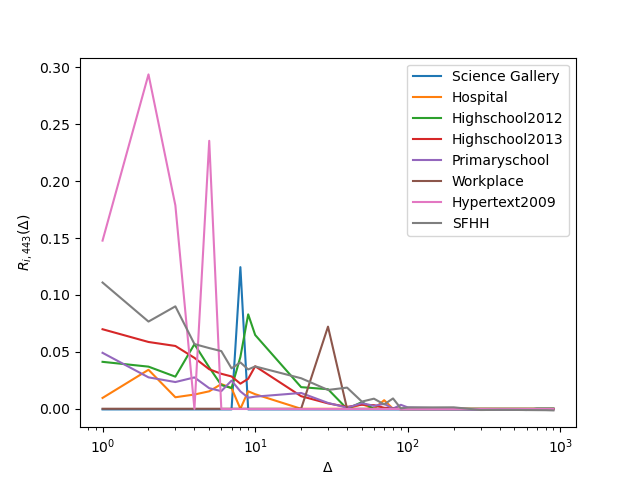}
        \caption{$\phi = 443$}
        \label{fig:pearsoncor_443}
    \end{subfigure}
    \caption{Average Pearson correlation coefficient $R_{i,\phi}(\Delta)$ for order $4$ hyperlinks in eight real-world physical contact networks as a function of time lag $\Delta$.}
    \label{fig:pearsoncor_rest4}
\end{figure*}

\begin{table}[h]
\begin{adjustbox}{center}
    \begin{tabular}{|c|c|c|c|c|c|c|c|c|c|c|}
        \hline
        Dataset & $c_{444}$ & $c_{421}$ & $c_{422}$ & $c_{431}$ & $c_{432}$ & $c_{433}$ & $c_{441}$ & $c_{442}$ & $c_{443}$ & $c_4$   \\
        \hline\hline
        Science Gallery & 0.23 & 0.00 & 0.01 & 0.00 & 0.01 & 0.07 & 0.00 & -0.01 & -0.13 & 0.00  \\  \hline
        Hospital & 0.11 & 0.00 & 0.00 & 0.00 & 0.00 & 0.03 & 0.00 & 0.00 & 0.00 & 0.00  \\  \hline
        Highschool2012 & 0.17 & 0.00 & 0.00 & 0.00 & 0.00 & 0.05 & 0.00 & 0.00 & 0.01 & 0.00 \\ \hline
        Highschool2013 & 0.36 & 0.00 & 0.00 & 0.00 & 0.00 & 0.06 & 0.00 & 0.00 & 0.01 & 0.00 \\ \hline
        Primaryschool & 0.09 & 0.00 & 0.00 & 0.00 & 0.00 & 0.04 & 0.00 & 0.00 & 0.02 & 0.00 \\ \hline
        Workplace & 0.00 & 0.00 & 0.00 & 0.00 & 0.00 & 0.00 & 0.00 & 0.00 & 0.00 & 0.00 \\ \hline
        Hypertext2009 & 0.00 & 0.00 & 0.00 & 0.00 & 0.00 & 0.03 & 0.00 & 0.02 & 0.00 & 0.00  \\ \hline
        SFHH Conference & 0.36 & 0.00 & 0.00 & 0.00 & 0.00 & 0.06 & 0.00 & 0.00 & 0.03 & 0.00 \\ \hline
    \end{tabular}
    \end{adjustbox}
\caption{Coefficients of the generalized model to predict order $4$ events, when $L=30$ and $\tau=5$. }
\label{tab:coeffs4}
\end{table}

\end{document}